\documentclass{jfm}

\usepackage{mathtools}
\usepackage[normalem]{ulem}
\usepackage[utf8]{inputenc}
\usepackage[T1]{fontenc}
\usepackage{mathptmx}
\usepackage{etoolbox}
\usepackage{epsfig}
\usepackage{epstopdf}
\usepackage{subcaption}
\usepackage{xcolor}
\usepackage{latexsym}
\usepackage{makecell}
\usepackage{hyperref}
\usepackage{xcolor}
\usepackage{booktabs}
\usepackage{multirow}
\usepackage{natbib}
\newcommand{\abs}[1]{\lvert#1\rvert}
\newcommand{\RN}[1]{%
  \textup{\lowercase\expandafter{\romannumeral#1}}%
}

\usepackage{todonotes}

\title{Dissolution-driven transport in a rotating horizontal cylinder}

\author{Subhankar Nandi\aff{1}, Jiten C. Kalita\aff{1}, Sanyasiraju VSS Yedida\aff{2} \and Satyajit Pramanik\aff{1} \corresp{\email{satyajitp@iitg.ac.in}}}

\affiliation{\aff{1}Department of Mathematics, Indian Institute of Technology Guwahati, Guwahati - 781039, India
\aff{2}Department of Mathematics, Indian Institute of Technology Madras, Chennai - 600036, India}

\begin{document}
\maketitle

\begin{abstract}
We study the combined effects of natural convection and rotation on the dissolution of a solute in a solvent-filled circular cylinder. The density of the fluid increases with the increasing concentration of the dissolved solute, and we model this using the Oberbeck-Boussinesq approximation. The underlying moving-boundary problem has been modelled by combining the Navier-Stokes equations with the advection-diffusion equation and a Stefan condition for the evolving solute-fluid interface. We use highly resolved numerical simulations to investigate the flow regimes, dissolution rates, and mixing of the dissolved solute for $Sc = 1$, $Ra \in [10^5, 10^8]$ and $\Omega \in [0, 2.5]$. In the absence of rotation and buoyancy, the distance of the interface from its initial position follows a square root relationship with time ($r_d \propto \sqrt{t}$), which ceases to exist at a later time due to the finite-size effect of the liquid domain. We then explore the rotation parameter, considering a range of rotation frequency -- from smaller to larger, relative to the inverse of the buoyancy-induced timescale -- and Rayleigh number. We show that the area of the dissolved solute varies nonlinearly with time depending on $Ra$ and $\Omega$. The symmetry breaking of the interface is best described in terms of $Ra/\Omega^2$. 
\end{abstract}

\begin{keywords}
Authors should not enter keywords on the manuscript, as these must be chosen by the author during the online submission process and will then be added during the typesetting process (see \href{https://www.cambridge.org/core/services/aop-file-manager/file/61436b61ff7f3cfab749ce3a/JFM-Keywords-Sept-2021.pdf.}{Keyword PDF} for the full list).  Other classifications will be added at the same time.
\end{keywords}


\section{Introduction}
The moving boundary problem \citep{crank1984, Pozrikidis_1997, Fokas_Kalimeris_2017} plays a pivotal role in understanding various natural processes and engineering applications where boundaries between different phases of matter or regions evolve. A prominent example of this is the Stefan problem, the simplest form \citep{VOLLER, MITCHELL2009, NANDI1} of which involves a partial differential equation for energy transport and a dynamic boundary. For example, in heat conduction problems, the phase change between solid and liquid phases creates a boundary that moves as the system evolves. The complexity of these problems arises from the need to model both the evolution of the boundary and the physical processes within the domain, such as heat diffusion or mass transfer, which are often nonlinear and coupled with the dynamics of the boundary itself. While this provides a simple framework for modeling phase changes, real-world systems are often more complex, requiring modifications to account for factors such as fluid movement or external disturbances. In many practical cases \citep{Stevens, stern1975ocean, mcphee2008air}, the involved phase-change material does not remain stationary but instead experiences convection due to energy gradients, buoyancy effects, or external forces. The phase-change in this case is not only driven by the energy diffusion but also by the convective flow within the liquid. This presents often more significant challenges both in mathematical modeling and physical understanding of the system, as it demands coupling of the flow fields with the energy transport mechanisms governed by convection-diffusion type equations, alongside the evolving boundary. One particular class of such problems arises in the context of dissolution processes \citep{Ristroph_2018, WykesD, pegler2020, Miao, wells_worster_2011, Nandi4}, where a solute material dissolves into a surrounding fluid.

Over the years, convective dissolution phenomena have been studied across various contexts in the literature, focusing on different aspects such as the shape evolution of dissolving materials \citep{Huang2022morphological, pegler2020, pegler2021}, employment of scaling laws \citep{Huang_2015, Dietrich_2016, cohen2020buoyancy}, flow regime analysis with their impact on the process \citep{HUANG, Nandi4, Timothy_1996} and so on. Despite their apparent diversity across different systems, the physical mechanisms driving these dissolution processes are governed by common underlying principles. In many cases, the dissolution of a solute into a surrounding fluid yields local variations in solute concentration, which subsequently change the fluid’s density. This density variation is the primary mechanism for the formation of convective flow as the system seeks to restore mass equilibrium. The buoyant forces caused by the density gradients lead to natural convection, which significantly affects the dissolution rate and the overall shape of the dissolving boundary. In many practical situations, in addition to these internal, naturally driven forces, external factors such as temperature gradients, shear forces, pressure gradients, or external flow fields can also play a significant role in influencing the convective dissolution process. For instance, temperature gradients \citep{naviaux2019temperature, olivella1996porosity}, or pressure changes \citep{ren2024dissolution, tasaka1990relaxation} can modify the dissolution dynamics by affecting both the fluid's properties and the solubility of the material. Similarly, rotation of the fluid or the presence of mechanical stirring generates shear forces \citep{wallin1989mass, ashokbhai2024drug, khoury1988dissolution} within the fluid, which can enhance mixing and disturb the formation of concentration gradients near the dissolving material. However, the combined effect of these factors, along with the internal forces, determines the overall efficiency of the dissolution process. 

The process driven by the interplay between internal forces with rotational factors is particularly relevant in various industrial applications such as drug dissolution, mixing processes, and solute extraction from materials. Most of the studies found in the existing literature on this area are based on experimental results. For example, one may refer to the study on the dissolution of limestone in a rotating cylinder \citep{wallin1989mass}, dissolution of hydrocortisone alcohol and acetate in rotating fluids \citep{khoury1988dissolution}, the rotating cylinder electrode \citep{gabe1974rotating}, and so on. In many of these studies, the typical setup consists of a dissolvable solute placed in a rotating vertical cylindrical container filled with solvent. On the other hand, in the current study, though we focus on understanding a similar dissolution process,  apart from the solute being in the shape of a circular cylinder, such as a candle or rod, the rotating cylindrical container is horizontal, rather than vertical. A previous study on the dissolution in a similar configuration but without rotation was conducted by ~\citet{Nandi4}. They mainly focused on the dissolution rate and shape evolution under various parameters like fluid-solute volumetric ratio, Stefan Number $(St)$, solutal Rayleigh number $(Ra)$ and Schmidt number $(Sc)$, etc. However, this study specifically aims to examine the contribution of rotational forces to the dissolution process while keeping all other involved parameters constant. Our objective is to observe how the rate of dissolution differs with the addition of rotational force, along with the corresponding flow behaviors and shape dynamics of the solute. Additionally, the study attempts to predict dissolution time at various rotation speeds and explores the relationship between the percentage of dissolution and time. It also focuses on providing a rough/estimated idea of shape dynamics by introducing a modified Rayleigh number $(Ra_{\Omega})$ that accounts for both gravitational and rotational effects. In the process, it also aims to provide a complete numerical framework for predicting the dissolution dynamics in rotating systems, which was lacking in the existing literature. 

Mathematically, modeling dissolution (phase-change) systems chosen for the present configuration is fraught with unique challenges due to the dynamic nature of the interface and the additional mechanical rotational factors introduced by the rotation of the cylinder. Traditional numerical techniques often fail to adequately address such complex, time-dependent, and multi-dimensional systems. This study adopts a stable and accurate boundary-fitted grid-based method \citep{Nandi3}, which is capable of solving the coupled non-linear governing equations while accurately capturing the transient interface without the need for supplementary tools such as level-set, phase field, immersed boundary methods, etc. Here, at each time step of computation, the physical domain is transformed into a fixed rectangular domain using a boundary-fitted transformation. The transformed energy transport and the flow-field equations are then solved on the computational grid using an unconditionally stable alternating direction implicit (ADI) scheme, while the boundary update equations are solved using a total variation diminishing (TVD) Runge–Kutta (RK) method.  Before applying this numerical approach to the problem under consideration, it is validated by solving a benchmark problem of natural convective heat transfer in a counter-rotating cylindrical annulus \citep{abu2007combined} but without phase transition using the same scheme. 

The outline of the paper is as follows. Section \ref{sec:math_model} details the mathematical formulation of the problem, including the governing equations and boundary conditions. Section \ref{sec:numerical_method} discusses the numerical methods used for solving the problem, while Section \ref{sec:diss_mix} presents and analyzes the results of our simulations. Finally, in Section \ref{sec:discussion}, we conclude by summarizing our findings and suggesting potential directions for future research.

\section{Mathematical Model} \label{sec:math_model}

\begin{figure*}
 \begin{center}
 \includegraphics[width=\textwidth]{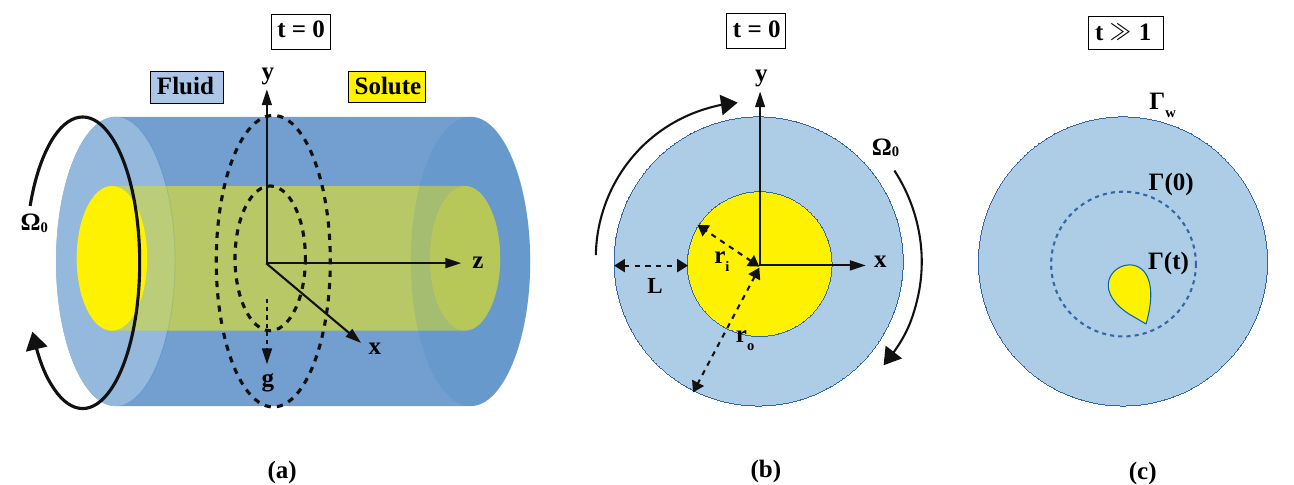}
  \caption{Schematic representation of the problem under consideration. (a) An infinitely long rotating horizontal cylinder filled with an incompressible solvent (blue) in which a solute (yellow) in the shape of a concentric circular cylinder is placed initially (at time $t = 0$), and (b) its $xy-$cross-section showing the two-dimensional nature of the problem under investigation in this study. (c) Location and geometry of the solute-solvent interface $\Gamma(t)$ at a later time ($t \gg 1$) along with its initial position, $\Gamma(0)$ (dashed line), and the cylinder wall, $\Gamma_w$.} \label{fig:schematic}
\end{center} 
\end{figure*}

Consider the dissolution 
of a cylindrical solvable substance (solid/solute) in a solvent (fluid) filling an infinitely long horizontal 
concentric cylinder
rotating in the clockwise direction about its axis (see figure \ref{fig:schematic}). 
As the solid dissolves, the fluid/solid boundary (interface) recedes, dissolved solute mixes with the solvent and forms a solution (solvent + dissolved solute). 
It is assumed that the solute is sufficiently solid so that it dissolves uniformly in the solvent without fragmenting under the influence of the developed flow. In general, the fragmentation of a dissolving substance depends on several factors, including mechanical stress from high fluid velocities, chemical instability or reactivity in the solvent, solubility limits, thermal effects, and so on. 
Solute dissolution leads to a density gradient in the fluid, leading to a convective flow. 

Further, we assume that the cylinder experiences a clockwise rotation relative to the solute at an angular velocity $\Omega_0$ about its horizontal axis. 
Although the rotation of the cylinder enhances the convective flow, the assumptions of the flow remain valid up to a certain rotation speed till the solubility limit is maintained. 
It is worth mentioning that due to the mechanically induced forced convection resulting from cylinder rotation, the solute dissolution is driven by the combined effects of natural and forced convection. 


Assuming that the flow remains invariant along the axis of the cylinder, the problem can be modelled mathematically in two space dimensions. We consider a cross-section of the cylinder perpendicular to its axis such that the liquid region is contained between the cylinder and the interface (see figure \ref{fig:schematic}(b) for the initial time and figure \ref{fig:schematic}(c) for a later time). 
We restrict our study to the fluid region, and the boundary concentration of the solute is assumed to be equal to the phase change concentration at which the solution saturates. Solute dissolution may generate latent heat due to the interactions between the solute and solvent molecules. Depending on the nature of the solute-solvent interactions, it may either absorb or release latent heat. However, the effects of the latent heat generated and heat arising due to mixing during the dissolution are extremely negligible on the flow structure, and hence they are neglected here. Thus, the density of the solution depends solely on the concentration of the dissolved solute. 

The fluid region, $\Sigma_{l}(t)\subset \mathbb{R}^2$, is bounded by the rigid wall of the cylinder, $\Gamma_w$ (outer boundary), and the interface, $\Gamma(t)$ (inner boundary). At $t = 0$, the fluid region corresponds to an annular region (see figure \ref{fig:schematic}(b)), and as time progresses, its shape is determined by the evolving interface $\Gamma(t)$ (see figure \ref{fig:schematic}(c)) governed by the Stefan condition. 
The liquid is assumed to be incompressible, and the density variation is modelled using the Oberbeck-Boussinesq approximation. The mass balance of the dissolved solute concentration ($c$) is described in terms of an advection-diffusion equation. Thus, the resulting governing equations read as 
\begin{align}
& \vec{\nabla} \cdot \vec{u} = 0,  \qquad \mbox{in} \qquad \Sigma_{l}(t) \times (0, \infty),\label{eq:nvs_a} \\
&\frac{\partial \vec{u}}{\partial t} + (\vec{u}\cdot \vec{\nabla}) \vec{u}  = - \frac{1}{\rho_0} \vec{\nabla} p + \nu \nabla^2 \vec{u} + \hat{g} \beta_c c,  \qquad \mbox{in} \qquad \Sigma_{l}(t) \times (0, \infty), \label{eq:nvs_b} \\ 
& \frac{\partial c}{\partial t} + \vec{u}\cdot \vec{\nabla} c  = D \nabla^2 c,  \qquad \mbox{in} \qquad \Sigma_{l}(t) \times (0, \infty), \label{eq:cde_dimensional}
\end{align}
where $p$ represents the pressure distribution, $\hat{g}$ denotes the gravitational acceleration, $\rho_0 = \rho(c = 0)$ is the density of the solvent, $\beta_c$ the solutal mass expansion coefficient, $c_{\Gamma}$ the concentration at the interface, $\nu$ the kinematic viscosity and $D$  the mass diffusivity of the solute in the liquid. The last term on the right side of equation \eqref{eq:nvs_b} indicates the buoyancy force acting against the force of gravity. Here, we consider that the gravity acts vertically downward, i.e., $\hat{g} = -g\hat{j}$, where $\hat{j}$ denotes the unit vector in the $y$-direction. 

\subsection{Initial and Boundary conditions} \label{subsec:IC_BCs}

The mathematical description of the problem is concluded by prescribing initial and boundary conditions. Initially, the fluid region is filled with quiescent solvent, mathematically 
\begin{equation}
    \label{eq:IC_dimensional}
    \vec{u}(\vec{x}, 0) = \vec{0}, \quad c(\vec{x}, 0) = 0 \qquad \mbox{in} \;\;\; \Sigma_l(0). 
\end{equation}

Fluid velocity at the boundaries satisfy 
\begin{equation} 
\label{eq:boundary_u}
    \vec{u} = V \hat{n}^\Gamma = (V_x, V_y )  \quad \mbox{ on } \;\;\; \Gamma(t), \quad \quad \vec{u} = (\Omega_0 y, -\Omega_0 x ) \quad \mbox{ on } \;\;\; \Gamma_w.
\end{equation}
where $V_x$, $V_y$ are the horizontal and vertical components of the interface velocity, respectively, and $\hat{n}^\Gamma = (n_x^\Gamma, n_y^\Gamma)$ is the unit normal to the interface into the fluid region. Although the fluid velocity at the interface is assumed to be the same as the interface velocity, one can also use $V = 0$ \citep{HUANG}. This assumption is relevant because the interface velocity is typically very small ($\sim \mathbf{O}(10^{-5})$) in dissolution problems. On the other hand, the fluid velocity at the wall $\Gamma_w$ is specified based on a fixed clockwise rotation speed $\Omega_0 (\ge 0)$. This indicates that the wall is treated as rotating with a constant angular velocity, without any acceleration or oscillation. 

At the interface $\Gamma(t)$, concentration equals the saturation concentration ($c_{sat}$) of the solution, 
\begin{equation} 
\label{eq:BC_conc_interface_dim}
c_{\Gamma} := c(\vec{x},t) = c_{sat} \quad \quad \mbox{ on } \;\;\; \Gamma(t). 
\end{equation}
As the solute dissolves, the interface recedes as a necessary consequence of mass conservation, described by the flux balance 
\citep[][and references therein]{wells_worster_2011, pegler2021}
\begin{equation} 
    \label{eq:stefan_dim}
    \rho_s \left(c_s - c_{\Gamma} \right) V = - \rho_0 D \vec{\nabla}c \cdot \vec{n}^\Gamma \quad \quad \mbox{ on } \;\;\; \Gamma(t),
\end{equation}
where $\rho_s$ is the density of the solid, $c_s$ is the concentration of the undissolved solute, $V$ is the normal velocity of the interface.
The condition \eqref{eq:stefan_dim} indicates that the solute dissolution rate/speed depends on the concentration flux at the interface as well as the mass diffusivity of the solute in the solvent. Interestingly, in nature, the mass diffusivity of most solute materials is relatively low. As a result, we can expect the dissolution process to be slower than other phase-change processes, such as melting or solidification. However, the current investigation does not focus on any specific solute or solvent. At the outer boundary $\Gamma_w$, we assume a Neumann condition 
\begin{equation}
    \label{eq:BC_conc_wall_dim}
    -D \vec{\nabla}c \cdot \vec{n}^w = 0  \quad \mbox{ on } \;\;\; \Gamma_w
\end{equation}
representing no diffusive flux, wherein $\vec{n}^w$ represents a unit normal out of the fluid region. 

\subsection{Non-dimensionalization} \label{subsec:non-dim}

Introducing the annular gap width $L$ as the characteristic length scale, we define the characteristic velocity as
\begin{equation}\label{eq:refU}
u_c = \left( g \beta_c L \Delta c \right)^{\frac{1}{2}}, 
\end{equation}
which is analogous to the free-fall velocity commonly used in natural convection, obtained by balancing the buoyancy and convective terms appearing in the momentum equation. Using this velocity scale, we render the non-dimensional variables as
\begin{equation}
\label{eq:nondim_variable}
\vec{x}^\prime = \frac{\vec{x}}{L},\quad t^\prime = \frac{t u_c}{L},\quad \vec{u}^\prime = \frac{\vec{u}}{u_c},
\quad p^\prime = \frac{p}{\rho u_{c}^2}, \quad c^\prime = \frac{c}{\Delta c},  \quad \mbox{ where } \quad \Delta c = \max_{t = 0}{(\abs{c - c_{\Gamma}})}. 
\end{equation}
The corresponding dimensionless equations read (after dropping the prime symbols )
\begin{align}
& \vec{\nabla} \cdot \vec{u} = 0,  \quad \quad \mbox{ in } \Sigma_{l}(t) \times (0, \infty),\label{eq:flowconti_nondim} \\
&\frac{\partial \vec{u}}{\partial t} + (\vec{u}\cdot \vec{\nabla}) \vec{u}  = -  \vec{\nabla} p + \frac{1}{Re} \nabla^2 \vec{u} - c \hat{j}, \quad \quad \mbox{ in } \Sigma_{l}(t) \times (0, \infty), \label{eq:nvs_nondim} \\
&\frac{\partial c}{\partial t} + \vec{u}\cdot \vec{\nabla} c = \frac{1}{Pe} \nabla^2 c, \quad \quad \mbox{ in } \Sigma_{l}(t) \times (0, \infty). \label{eq:con_nondim} 
\end{align}
Equations \eqref{eq:flowconti_nondim}-\eqref{eq:con_nondim} are associated with the initial conditions
\begin{equation}
    \label{eq:IC_nondim}
    \vec{u}(\vec{x}, 0) = \vec{0}, \quad c(\vec{x}, 0) = 0 \qquad \mbox{in} \;\;\; \Sigma_l(0), 
\end{equation}
boundary conditions 
\begin{align}
    & c = 1  \quad \mbox{ on } \;\;\; \Gamma(t), \qquad \qquad \frac{\partial c}{\partial n^w} = 0  \quad \mbox{ on } \Gamma_w, \label{eq:boundary_c} \\
    & \vec{u} = V \hat{n^w} \quad \mbox{ on } \;\;\; \Gamma(t), \qquad \qquad \vec{u} = (\Omega y, -\Omega x ) \quad \;\;\; \mbox{ on } \Gamma_w, \label{eq:boundary_u_nondim} 
\end{align}
and the Stefan condition 
\begin{align}
    & V = \frac{St}{Pe} \frac{\partial c}{\partial n^\Gamma} \quad \quad \mbox{ on } \;\;\; \Gamma(t). \label{eq:stefan_nondim}
\end{align}

A close look into equations \eqref{eq:flowconti_nondim}-\eqref{eq:stefan_nondim} yields that the convective dissolution problem can be studied in term of four dimensionless parameter --- P\'eclet number ($Pe$), Reynolds number ($Re$) (or, two derived parameters -- solutal Rayleigh number ($Ra = Re \cdot Pe$), Schmidt number ($Sc = Pe/Re$)), Stefan number ($St$) and $\Omega$. These parameters are defined as
\begin{align}
\label{eq:nondim_parameter}
& Re = \frac{u_c L}{\nu}, \;\; Pe = \frac{u_c L}{D}, \;\; 
Ra = 
\frac{g \beta_c \Delta c L^3}{\nu D}, \;\;  
Sc = 
\frac{\nu}{D}, \;\;
 St = \frac{\rho_0 \Delta c}{\rho_s \Delta c_s}, \;\; \Omega = \frac{\Omega_0 L}{u_c}, 
\end{align}
where $\Delta c_s = c_s - c_\Gamma$.

\subsection{Stream function-vorticity formulation} \label{subsec:stream_vorticity}

The stream function-vorticity formulation of the Navier-Stokes equations in a two-dimensional domain facilitates a more comprehensive and intuitive understanding of flow mechanisms than the conventional representation using primitive variables ($\vec{u}, p$). In a two-dimensional Cartesian framework, the relationships between the stream function $(\psi)$ and vorticity $(\omega)$ are defined as follows:
\begin{equation}
(u,v) = \left( \frac{\partial \psi}{\partial y},  - \frac{\partial \psi}{\partial x} \right), \quad \omega = \frac{\partial v}{\partial x} - \frac{\partial u}{\partial y}, \label{eq.psi_omg.relation}
\end{equation}
\noindent where $ u,v $ denote the velocity components of the fluid in $x,y$ directions, respectively. Using these relations, the governing flow system becomes
\begin{align}
&\frac{\partial^2 \psi}{\partial x^2} + \frac{\partial^2 \psi}{\partial y^2} = - \omega, \quad \quad \mbox{ in } \Sigma_{l}(t) \times (0, \infty), \label{eq:psi_poisson} \\
& \frac{\partial \omega}{\partial t} + (\vec{u}\cdot \vec{\nabla}) \omega  =  \frac{1}{Re} \nabla^2 \omega - \frac{\partial c}{\partial x}, \quad \quad \mbox{ in } \Sigma_{l}(t) \times (0, \infty). \label{eq:omg_nvs}
\end{align}
and the corresponding boundary conditions for $\psi$ are 
\begin{align}
&\psi = 0, \quad \frac{\partial \psi}{\partial n^\Gamma} = V_x n_x^\Gamma + V_y n_y^\Gamma, \quad \quad \mbox{ on } \Gamma(t), \label{eq:boundary_psi1} \\
& \psi = 0, \quad \frac{\partial \psi}{\partial n^w} =  \Omega (x n_x^w + y n_y^w), \quad \quad \mbox{ on } \Gamma_w. \label{eq:boundary_psi2} 
\end{align}
To complete the system, boundary conditions for the vorticity $\omega$ are required to be computed from equation \eqref{eq:psi_poisson} utilizing conditions \eqref{eq:boundary_psi1} and \eqref{eq:boundary_psi2}. These are to be discussed/specified later.


\subsection{Gibbs-Thomson effect} \label{subsec:Gibbs_Thomson} 

The Stefan condition \eqref{eq:stefan_nondim} indicates that the interface velocity can be determined solely from the parameter ratio $St/Pe$ and the concentration flux into the fluid region at the interface. Notably, this movement/velocity may also be influenced by the shape of the interface surface. In the convex regions, where the local curvature exceeds the mean curvature, the interface moves more rapidly, enhancing the dissolution process. Conversely, in concave regions, the movement is steadier, resulting in a slower dissolution process. Although the initial circular interface of this model is smooth, it may develop irregularities in areas of both low and high curvature during dissolution. In such instances, it is necessary to equilibrate the interface motion between the concave and convex regions. The Gibbs-Thomson effect effectively addresses this requirement by establishing equilibrium in mass exchange between the solid and fluid regions due to interface effects. In this context, the interface velocity is adjusted as \citep{moore2013self, HUANG, PEREZ2005}
\begin{equation}
V = \frac{St}{Pe} \frac{\partial c}{\partial n} - \epsilon \left( \kappa - \frac{2 \pi}{S} \right) \quad \quad \mbox{ on } \;\;\; \Gamma, \label{eq:stefan_gibbs}
\end{equation}
where $\epsilon$ is smoothing term in the interface velocity to enhance numerical stability without affecting the physical dissolution process \citep{moore2013self, HUANG}.
Here, $\kappa$, $S$ represent the local curvature and total arc length of the interface, respectively.
When the interface is circular, the local curvature $\kappa$ is equivalent to the mean curvature given by $2\pi/S$. As a result, no adjustment to the interface velocity is required for a steadily advancing circular interface.

\section{Numerical Methods} \label{sec:numerical_method}

\subsection{Transformed model in computational domain} \label{subsec:computational_model}

\begin{figure}
  \begin{subfigure}{6.88 cm}
   \includegraphics[width=0.9\textwidth]{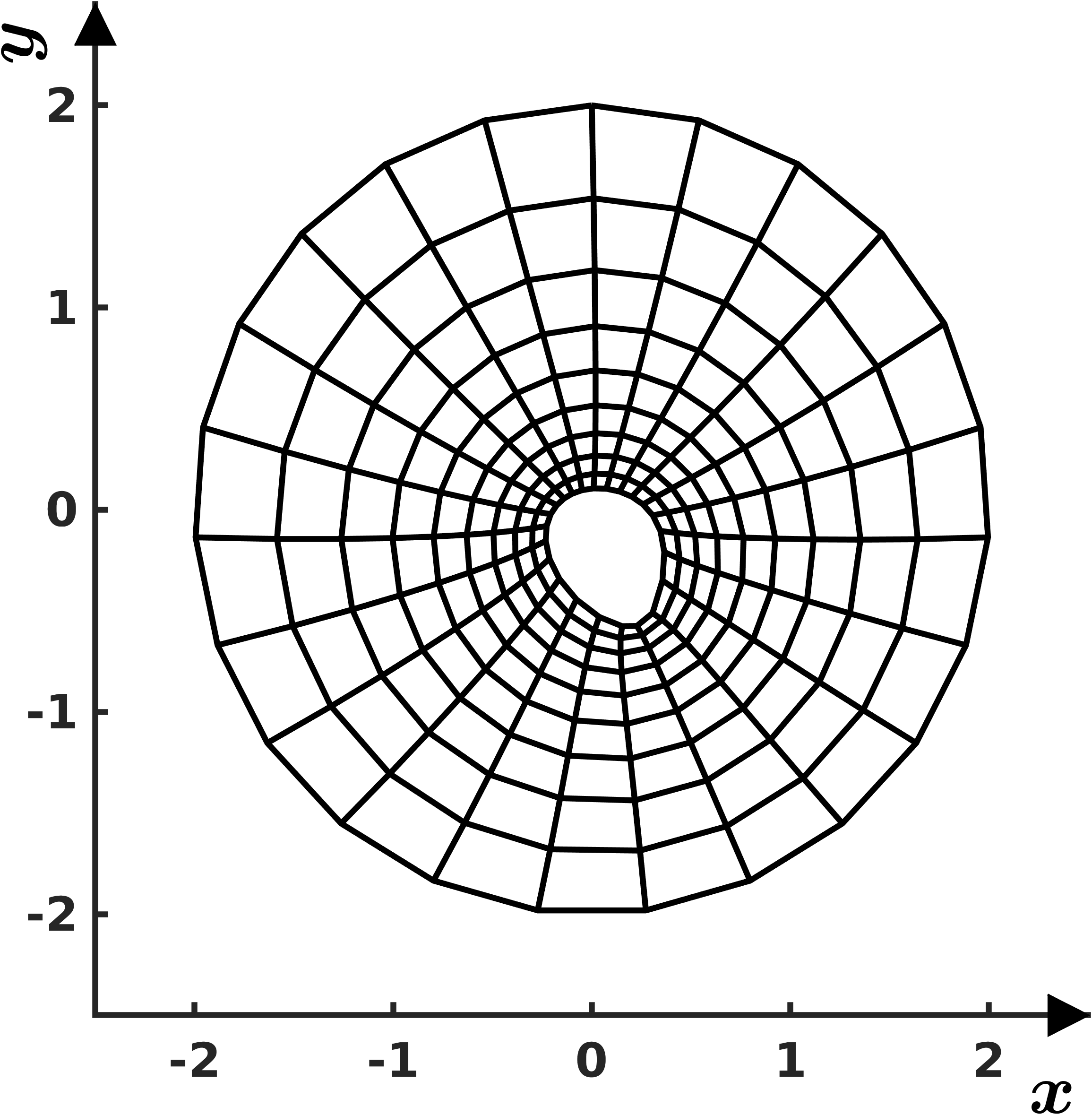}
    \caption{} 
  \end{subfigure}
  \begin{subfigure}{6.8 cm}
    \includegraphics[width=0.9\textwidth]{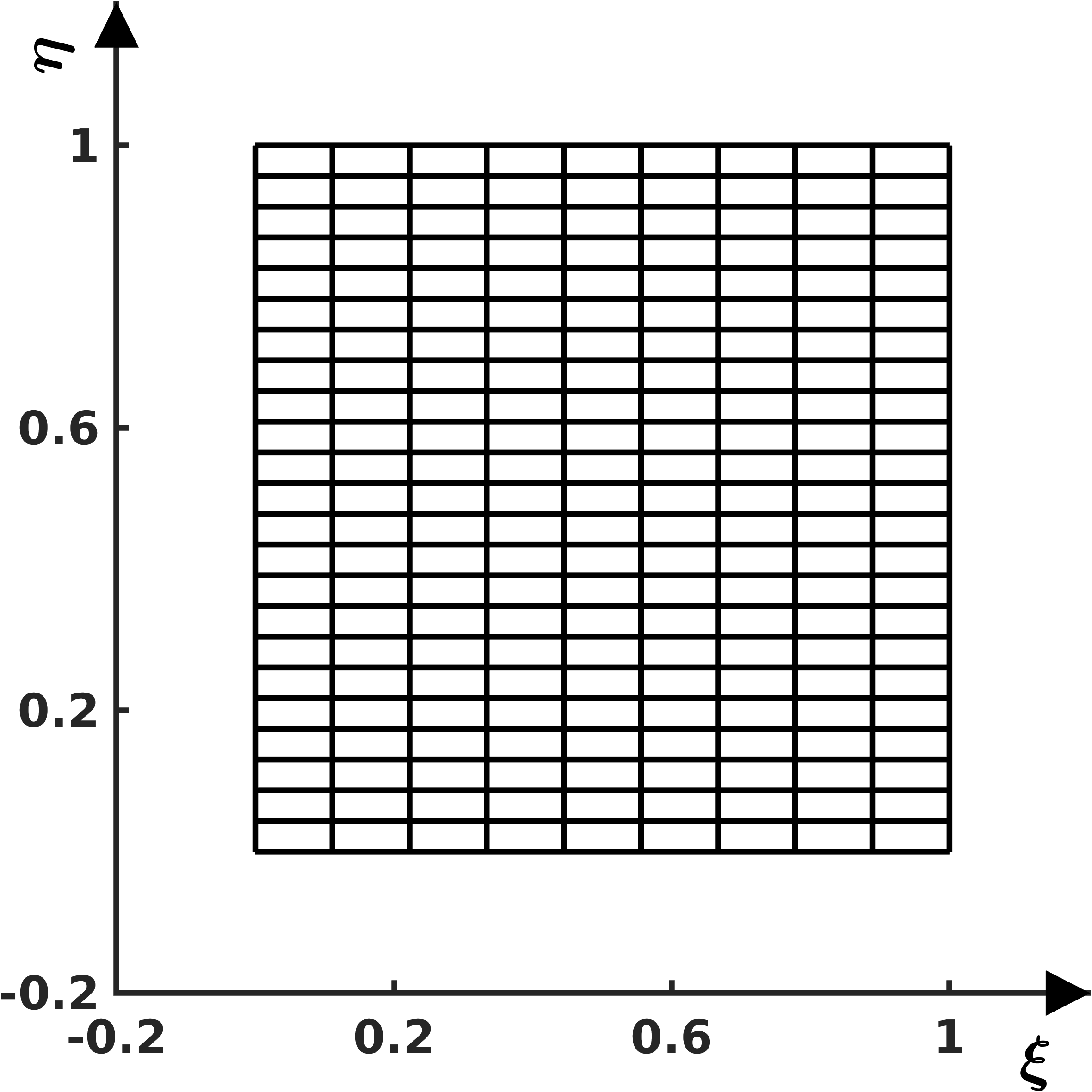}
    \caption{}
  \end{subfigure}
  \caption{Layout of a typical $10\times 24$ grid  in the (a)  physical $(x,y)$ domain and (b) computational $(\xi,\eta)$ domain.} \label{fig:sample_grid}
\end{figure}

The initial annular domain of the problem evolves due to the dynamic characteristics of the interface, which may lead to complex shapes. Therefore, in order to solve the problem on a uniform Cartesian grid numerically, the physical domain is transformed at every time step into a fixed unit square $([0,1]\times[0,1])$ using a boundary-fitted transformation (see figure \ref{fig:sample_grid}). This transformation is implemented so that the physical boundaries of the problem align with the boundary curves of the $(\xi,\eta)$-coordinates. In the present scenario, we map the inner boundary $\Gamma(t)$ (the interface) to the line $\xi = 0$ and the outer rotating wall $\Gamma_w$ to the line $\xi = 1$. The other two lines, $\eta = 0$ and $\eta = 1$, act as artificial boundaries, allowing the computational domain to be defined without resorting to the use of physical boundaries.

The corresponding Cartesian grid in $(\xi,\eta)$-domain is generated by solving the equations \citep[see][for further details]{THOMPSON, NANDI2}: 
\begin{align}
& \alpha x_{\xi \xi}  - 2\beta x_{\xi \eta} + \gamma x_{\eta \eta} + J^{2} \left(Px_{\xi} + Qx_{\eta} \right) = 0, \label{eq:grid_x} \\ 
& \alpha y_{\xi \xi}  - 2\beta y_{\xi \eta} + \gamma y_{\eta \eta} + J^{2} \left(Py_{\xi} + Qy_{\eta} \right) = 0, \label{eq:grid_y} 
\end{align}
with
\begin{align}
&\alpha = x_{\eta}^{2} + y_{\eta}^{2}, \quad \beta = x_{\xi} x_{\eta} + y_{\xi} y_{\eta} ,  \quad \gamma =  x_{\xi}^{2} + y_{\xi}^{2}, \quad J =  x_{\xi} y_{\eta} - y_{\xi} x_{\eta}. 
\end{align} 
Here, the functions $P$ and $Q$ are used to control the spacing between grid lines. To avoid the interpolation at every time step of calculations, time derivative is calculated in the $(\xi, \eta)$-domain using the following relation
\begin{equation}
\label{eq:ex_timeDr}
\begin{split}
\Big(\frac{\partial g}{\partial t}\Big)_{x, y} & = \frac{\partial (x, y, g)}{\partial (\xi, \eta, t)}/\frac{\partial (x, y, t)}{\partial (\xi, \eta, t)} \\
 & = \Big(\frac{\partial g}{\partial t}\Big)_{\xi, \eta} - \frac{(g_{\xi}y_{\eta}  - g_{\eta}y_{\xi})}{J}  \Big(\frac{dx}{dt}\Big)_{\xi, \eta} + \frac{(g_{\xi}x_{\eta}  - g_{\eta}x_{\xi})}{J} \Big(\frac{dy}{dt}\Big)_{\xi, \eta} 
\end{split}
\end{equation}
where $g \in \{ c, \omega \}$.  

Using \eqref{eq:ex_timeDr}, governing equations for the concentration field $c$ in the computational plane reads as 
\begin{equation}
\label{eq:comp_c}
\Big(\frac{\partial c}{\partial t}\Big)_{\xi, \eta} = A c_{\xi \xi} + B c_{\xi \eta}  + C c_{\eta \eta}  + D c_{\xi} + E c_{\eta}
 - \left(u^{\xi}c_{\xi} + u^{\eta} c_{\eta} \right), 
\end{equation}
where
\begin{equation}
\begin{split}
& A = \frac{\alpha}{J^{2}Pe},\quad B = - \frac{2 \beta}{J^{2}Pe},\quad C = \frac{\gamma}{J^{2}Pe},\\
& D = \frac{P}{Pe} - \frac{1}{J} \left(x_{\eta}\frac{dy}{dt} -  y_{\eta} \frac{dx}{dt}\right), \quad  E = \frac{Q}{Pe} - \frac{1}{J} \left( y_{\xi} \frac{dx}{dt} -  x_{\xi} \frac{dy}{dt}\right).
\end{split}
\end{equation}
Similarly, the equation for the vorticity field can be expressed as
\begin{equation}
\label{eq:comp_vorticity}
\begin{split}
\Big(\frac{\partial \omega}{\partial t}\Big)_{\xi, \eta} =  A_1 \omega_{\xi \xi}  + B_1 \omega_{\xi \eta}  + C_1 \omega_{\eta \eta} + D_1 \omega_{\xi} + E_1 \omega_{\eta} - \left(u^{\xi} \omega_{\xi} + u^{\eta} \omega_{\eta} \right) + R, 
\end{split} 
\end{equation}
where
\begin{equation}
\begin{split}
& A_1 = \frac{\alpha}{J^{2}Re},\quad B_1 = - \frac{2 \beta}{J^{2}Re},\quad C_1 = \frac{\gamma}{J^{2}Re}, \quad R = - \frac{1}{J} \left( c_\xi y_\eta -  c_\eta y_\xi \right),\\
& D_1 = \frac{P}{Re} - \frac{1}{J} \left( x_{\eta} \frac{dy}{dt} -  y_{\eta} \frac{dx}{dt}\right), \quad E_1 = \frac{Q}{Re} - \frac{1}{J} \left( y_{\xi} \frac{dx}{dt} -  x_{\xi} \frac{dy}{dt} \right).
\end{split}
\end{equation}

Here, the contravariant fluid velocity components $u^{\xi}$ and $u^{\eta}$ in the $\xi$- and $\eta$- directions are expressed as follows:
\begin{equation}
\label{eq:comp_u_xi_eta}
u^{\xi} = \vec{u} \cdot \vec{\nabla} \xi = \frac{\psi_\eta}{J}, \quad u^{\eta} = \vec{u} \cdot \vec{\nabla} \eta = -\frac{\psi_\xi}{J},
\end{equation}
which satisfy the continuity equation \eqref{eq:flowconti_nondim}. Therefore, in computations, techniques such as upwinding or downwinding can be applied using these contravariant components instead of Cartesian ones. This approach facilitates an accurate numerical representation of fluid flow across different coordinate systems. The stream function and the interface normal velocity in $\xi\eta$-plane are described by 
\begin{align}
    & \alpha \psi_{\xi \xi} - 2\beta  \psi_{\xi \eta} + \gamma  \psi_{\eta \eta} + J^{2} P \psi_{\xi } + J^{2}Q \psi_{\eta} = - J^{2}\omega, \label{eq:comp_psi} \\
    & V = - \frac{St}{J Pe \sqrt{\alpha}}\left(\alpha c_{\xi} - \beta c_{\eta} \right) - \epsilon \left( \kappa - \frac{2 \pi}{S} \right), \quad \quad \mbox{ on } \xi =0 . \label{eq:comp_stefan}
\end{align}

At the artificial boundaries $\eta = 0$ and $\eta = 1$, the flow variables are 
periodic.
On the other hand, at the other boundaries ($\xi = 0, 1$), concentration $c$ and stream function $\psi$ are computed using \eqref{eq:boundary_c}, \eqref{eq:boundary_psi1}, and \eqref{eq:boundary_psi2}. Finally, the boundary conditions for $\omega$ at these boundaries are derived from the conditions for $\psi$ and equation \eqref{eq:comp_psi}. For example, at the boundary $\xi = 1$, equation \eqref{eq:boundary_psi2} yields, 
\begin{align}
& \psi_{\eta} = 0, \quad \psi_{\eta \eta} = 0,\\
& \psi_\xi = \frac{J \Omega}{\alpha} \left( xy_\eta - yx_\eta \right) = p_1 \quad \text{(say)}.
\end{align}
This leads to the equation for $\omega$ as follows:
\begin{equation}
\omega = -\frac{1}{J^2} \left( \alpha \psi_{\xi \xi} - 2\beta \psi_{\xi \eta} + J^2 P p_1 \right). \label{eq:comp_boundary_vort}
\end{equation}
A similar approach can be used to obtain the condition for $\omega$ at $\xi = 0$. It is important to note that the derivatives involved in the $\omega$ boundary condition are computed using a 5-point ghost cell technique to incorporate the specified lower-order non-zero derivative condition efficiently. 

\subsection{Numerical scheme implementation for flow and concentration} \label{subsec:numerical_scheme}

While implementing the numerical scheme, the spatial derivatives are discretized using standard second-order central difference approximations, while the convection terms are approximated using a third-order QUICK scheme \citep{leonard1995order, johnson1992equivalent}, as described below:
\begin{equation}\label{eq.upwind}
\left(u^{\xi} c_{\xi}\right)_{i} = \frac{u^{\xi}_{i} + \abs{u^{\xi}_{i}}}{2} \phi^{+}  + \frac{u^{\xi}_{i} - \abs{u^{\xi}_{i}}}{2} \phi^{-},
\end{equation}
where 
\begin{align}
& \phi^{+} = \frac{1}{6h} \left( \phi_{i-2} - 6\phi_{i-1} + 3\phi_{i} + 2\phi_{i+1}  \right), \\
& \phi^{-} = \frac{1}{6h} \left( -2\phi_{i-1} - 3\phi_{i} + 6\phi_{i+1} - \phi_{i+2}  \right).
\end{align}

To compute the stream function $\psi$, Poisson's equation \eqref{eq:comp_psi} is discretized with the boundary condition $\psi = 0$ at $\xi = 0, 1$. The resulting system of algebraic equations is solved using the Bi-conjugate gradient stabilized (BiCGStab) method with incomplete LU factorization as a preconditioner, up to a convergence tolerance of $10^{-10}$ of the residual vector. On the other hand, for the concentration $(c)$ and vorticity $(\omega)$ fields, the advection-diffusion equations \eqref{eq:comp_c} and \eqref{eq:comp_vorticity} are split in time direction using the ADI scheme \citep{in2009unconditional}
\begin{subequations}
\label{eq:ADI}
\begin{align}
&Y^{(0)} = c^{(n)} + \Delta tf(c^{(n)}), \label{eq:ADI_step1} \\ 
&Y^{(1)} = Y^{(0)} + \lambda \Delta t [ f_{1}(Y^{(1)}) - f_{1}(c^{(n)}) ], \label{eq:ADI_step2} \\
&Y^{(2)} = Y^{(1)} + \lambda\Delta t [ f_{2}(Y^{(2)}) - f_{2}(c^{(n)}) ], \label{eq:ADI_step3} \\
&Y^{(3)} = Y^{(0)} + \zeta \Delta t [f (Y^{(2)}) - f(c^{(n)}) ], \label{eq:ADI_step4} \\
&Y^{(4)} = Y^{(3)} + \lambda \Delta t [ f_{1}(Y^{(4)}) - f_{1}(Y^{(2)}) ], \label{eq:ADI_step5} \\
&c^{(n+1)} = Y^{(4)} + \lambda \Delta t [ f_{2}(Y^{(5)}) - f_{2}(Y^{(2)}) ], \label{eq:ADI_step6} 
\end{align}
\end{subequations}
where $f$ is defined as the sum of three functions -- $f_1$, which consists of derivatives in the $\xi$ direction; $f_2$, which involves derivatives in the $\eta$ direction; and $f_{12}$, which includes mixed derivatives and additional terms. The constant parameters $\lambda$ and $\zeta$ determining the stability of the ADI scheme are chosen as $\lambda \ge \frac{1}{2} + \frac{\sqrt{3}}{2}$ and $\zeta = \frac{1}{2}$, which are in accordance with the stability analysis \citep{NANDI2, in2009unconditional}. 

The ADI splitting \eqref{eq:ADI} involves six intermediate steps for computing the required variables for the next time step, comprising two explicit and four implicit steps. The internal variables $Y^{(0)}$ and $Y^{(3)}$ can be obtained directly from the first and fourth explicit steps, respectively. The other variables $Y^{(1)}, Y^{(2)}, Y^{(4)}$ and $Y^{(5)}$ are computed by solving the following four tri-diagonal systems 
\begin{subequations}
\label{eq:TDS}
\begin{align}
&(I - \lambda \Delta t S_{\xi})[Y^{(1)}] = [Y^{(0)}] - \lambda \Delta t S_{\xi}[c^{(n)}] + \lambda \Delta t [B_{1}], \label{eq:TDS1} \\
&(I - \lambda \Delta t S_{\eta})[Y^{(2)}] = [Y^{(1)}] - \lambda \Delta t S_{\eta}[c^{(n)}] + \lambda \Delta t [B_{2}], \label{eq:TDS2} \\
& (I - \lambda \Delta t S_{\xi})[Y^{(4)}] = [Y^{(3)}] - \lambda \Delta t S_{\xi}[Y^{(2)}] + \lambda \Delta t [B_{3}], \label{eq:TDS3} \\
& (I - \lambda \Delta t S_{\eta})[Y^{(5)}] = [Y^{(4)}] - \lambda \Delta t S_{\eta}[Y^{(2)}] + \lambda \Delta t [B_{4}], \label{eq:TDS4}
\end{align}
\end{subequations}
which are deduced from the remaining four implicit steps. Here, $S_\xi$ and $S_\eta$ are two tri-diagonal matrices derived from discretized equations involving functions $f_{1}(c)$ and $f_{2}(c)$. The column vectors $B_{l}$ (where $l = 1, \ldots, 4$ ) appearing in \eqref{eq:TDS1}-\eqref{eq:TDS4} represent the boundary conditions. The boundary conditions for the variables at the intermediate time steps are treated the same as the ones at the working time steps. 

It is worth noting that although there are four tri-diagonal systems, two distinct tri-diagonal matrices are sufficient to handle the computations for the required variables at a particular time step. The ADI scheme is preferred over other implicit or explicit methods, such as the Crank-Nicolson and total variation diminishing (TVD) Runge-Kutta schemes for solving the convection-diffusion equation for vorticity and concentration owing to the former's computational economy and unconditional stability. 

\subsection{Interface tracking} \label{subsec:comput_interface}

Recall from \S \ref{subsec:computational_model}, $\xi = 0$ corresponds to the interface. Therefore, the unit normal at the interface can be expressed as $\displaystyle \vec{n}^\Gamma = \left( \frac{y_\eta}{\sqrt{\alpha}}, -\frac{x_\eta}{\sqrt{\alpha}} \right)$. Thus, given the normal velocity $V$ from \eqref{eq:comp_stefan}, the interface in the $xy$-plane evolves according to  
\begin{equation}\label{eq:boundary_update_xy}
 \frac{dx}{dt} = \frac{V y_\eta}{\sqrt{\alpha}}, \qquad  
 \frac{dy}{dt} = -\frac{V x_\eta}{\sqrt{\alpha}}, 
\end{equation}
that can be expressed in a compact form, 
\begin{equation}
    \label{eq:boundary_update_xy_compact}
    \frac{d \mathbf{X}}{dt} = \mathbf{g}(\mathbf{X}, t),
\end{equation}
where $\mathbf{X} = (x, y)^\top$, and $\displaystyle \mathbf{g} = \left( V y_\eta/\sqrt{\alpha}, -V x_\eta/\sqrt{\alpha} \right)^\top$. 
Equations \eqref{eq:boundary_update_xy_compact} are solved using the third-order TVD Runge--Kutta scheme 
\begin{subequations}
\label{eq:TVD_rk}
\begin{align}
    & \mathbf{X}^{(1)} = \mathbf{X}^{(n)} + \Delta t \mathbf{g}(\mathbf{X}^{(n)}, t), \\
    & \mathbf{X}^{(2)} = \frac{3}{4} \mathbf{X}^{(n)} + \frac{1}{4} X^{(1)} + \frac{\Delta t}{4} \mathbf{g}(\mathbf{X}^{(1)}, t), \\
    & \mathbf{X}^{(n+1)} = \frac{1}{3} \mathbf{X}^{(n)} + \frac{2}{3} \mathbf{X}^{(2)} + \frac{2\Delta t}{3} \mathbf{g}(\mathbf{X}^{(2)}, t).
\end{align}
\end{subequations}
Here, we have used the TVD RK method instead of the previously used ADI scheme as the ODE system \eqref{eq:boundary_update_xy} involves only unidirectional derivative terms along the $\eta$-direction. However, it is worth noting that a first-order explicit scheme can also be used for solving these boundary update equations rather than higher-order methods, as its effects on the accuracy of the numerical solutions are negligible \citep{udaykumar1999computation}. 

\subsection{Overview of computational framework} \label{subsec:computational_framework}

Numerical simulations at a given time step consist of several key steps starting with the grid generation using equations \eqref{eq:grid_x}-\eqref{eq:grid_y}. Next, the stream function $\psi$ is calculated from equation \eqref{eq:comp_psi} based on the known values of vorticity $\omega$. Following this, the fluid velocity components $u^{\xi}$ and $u^{\eta}$ are derived from equation \eqref{eq:comp_u_xi_eta} using the stream function. The boundary conditions for $\omega$ are then set according to equation \eqref{eq:comp_boundary_vort}. Afterward, the concentration $c$ and vorticity $\omega$ are solved separately using the ADI scheme outlined in equations \eqref{eq:ADI}. Once these values are obtained, the moving interface is updated using equation \eqref{eq:boundary_update_xy} with the computed concentration field $c$, and the grid is regenerated to reflect the new interface location. This cycle of computation is repeated until convergence is achieved. The numerical solution is assumed to converge when the infinity norm of the absolute errors of the variables $\omega, c, x^\Gamma, y^\Gamma$ satisfies the specified tolerance of $10^{-8}$. Once the process converges, the flow field, concentration field, and interface are all stored at the current time step. 

The code validation and grid independence studies for our computations are presented in Appendices \ref{app:validation} and \ref{app:grid_independence}, respectively. The CPU time taken to complete one cycle of computation within the convergence loop is approximately 0.6 seconds for $50 \times 300$ spatial grid points using MATLAB 2022b in an Intel Xeon(R) Gold 6326 processor with a clock speed of 2.90 GHz and 64 GB of RAM.

\section{Dissolution and mixing} \label{sec:diss_mix}

Our primary objective is to study the combined effects of rotation and convection on dissolution and mixing. Throughout this paper, we fix $St/Pe = 3 \times 10^{-3}$, $Sc = 1$,  $\epsilon = 5 \times 10^{-4}$, $r_i/L = 1$ and $r_o/L = 2$. Thus, effects of buoyancy force ($Ra$) and rotation ($\Omega$) are studied by varying $Ra \in [10^5, 10^8]$ and $\Omega \in [0, 2.5]$.

\subsection{Qualitative impacts of rotation and buoyancy on flow and solute dissolution} \label{subsec:qualitative_flow_dissolution}

\begin{figure}
 \begin{center}
 \includegraphics[width=0.95\textwidth]{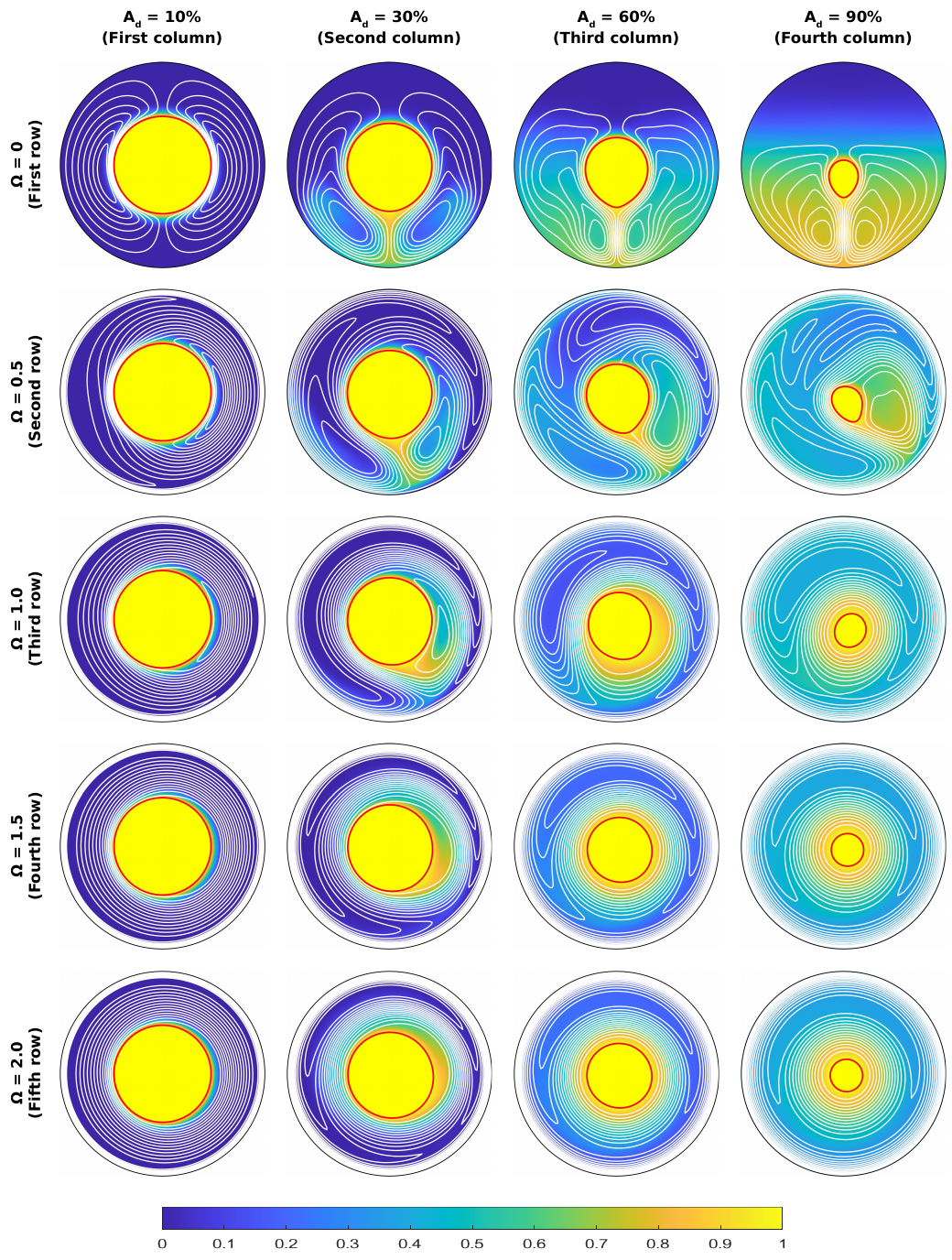}
  \caption{Streamlines (white contours) overlaying the concentration contours of the dissolved solute at $A_d(t) = 10\%$, $30\%$, $60\%$, and $90\%$ (left to right) for $Ra = 10^5$ and $\Omega =$ 0 to 2 with an increment of 0.5 (top to bottom). The corresponding colour map illustrates the concentration variations in the domain at these instances.} 
  \label{fig:flow_fixRa}
\end{center} 
\end{figure}

In this section, we discuss the spatio-temporal dynamics of the dissolved solute concentration and the flow field for different values of $Ra$ and $\Omega$. Figure \ref{fig:flow_fixRa} depicts concentration distribution overlaid by streamlines at four instances corresponding to $A_d(t) = 10\%, 30\%, 60\%, and 90\%$ (where $A_d$ corresponds to the dissolved area, defined in \eqref{eq:Ad}) for $Ra = 10^5$ and $\Omega = 0$ to 2 with an increment of 0.5. At the early stages ($A_d(t)$ approximately up to 10\%), irrespective of the rotation speed considered here, dissolution and hence the phase-change process is driven by diffusion only (see first column of figure \ref{fig:flow_fixRa}). Thus, the dissolved solute is localized near the interface, and its concentration reduces as we move away from the interface and eventually decays to zero near the wall. However, rotation breaks the left-right symmetry about the vertical axis $x = 0$, as evident from the first column of figure \ref{fig:flow_fixRa}. 

In the absence of rotation, the primary vortices, which are observed initially on either side of the solute along the horizontal central line, gradually move toward the bottom of the solute over time and remain there till the dissolution process is complete. On the other hand, with the rotation speed of $0.5$, a primary vortex forms on the right side of the solute's surface and eventually settles there even when dissolution reaches the $90\%$ stage. However, at the rotation speeds of $1,\;1.5$ and $2$, this vortex ceases to exist after certain stages of dissolution. 

Figure \ref{fig:flow_fixRa} also shows that, during dissolution with no rotation, the interplay between diffusion and buoyancy-driven convection aids in developing a laminar velocity boundary layer near the interface except at the topmost part. Rotation induces an additional laminar boundary layer near the cylinder wall. 
The boundary layer near the interface forms due to the combined effects of buoyancy and centripetal forces. At low rotational speeds, this layer is accompanied by a primary vortex to the right of the interface. As the rotation speed increases, the vortex is pushed away from the interface and eventually disappears. 
As the cylinder rotates, the fluid in immediate contact with the wall is dragged along, causing the fluid at the surface to move faster than the fluid away from the surface, which leads to the development of an additional boundary layer. An increasing rotation speed induces more shear and larger velocity gradients near the wall, affecting the nature of the boundary layer thickness. 
 
Another important observation is that the flow patterns at rotation speeds greater than $1$ begin to resemble concentric circular layers around both the solute and the cylindrical wall after a certain amount of dissolution. As the rotation speed increases, the gap between these layers becomes smaller. The flow also loses its closed trajectories and tends to become more uniform and structured. 

\begin{figure}
 \begin{center}
 \includegraphics[width=\textwidth]{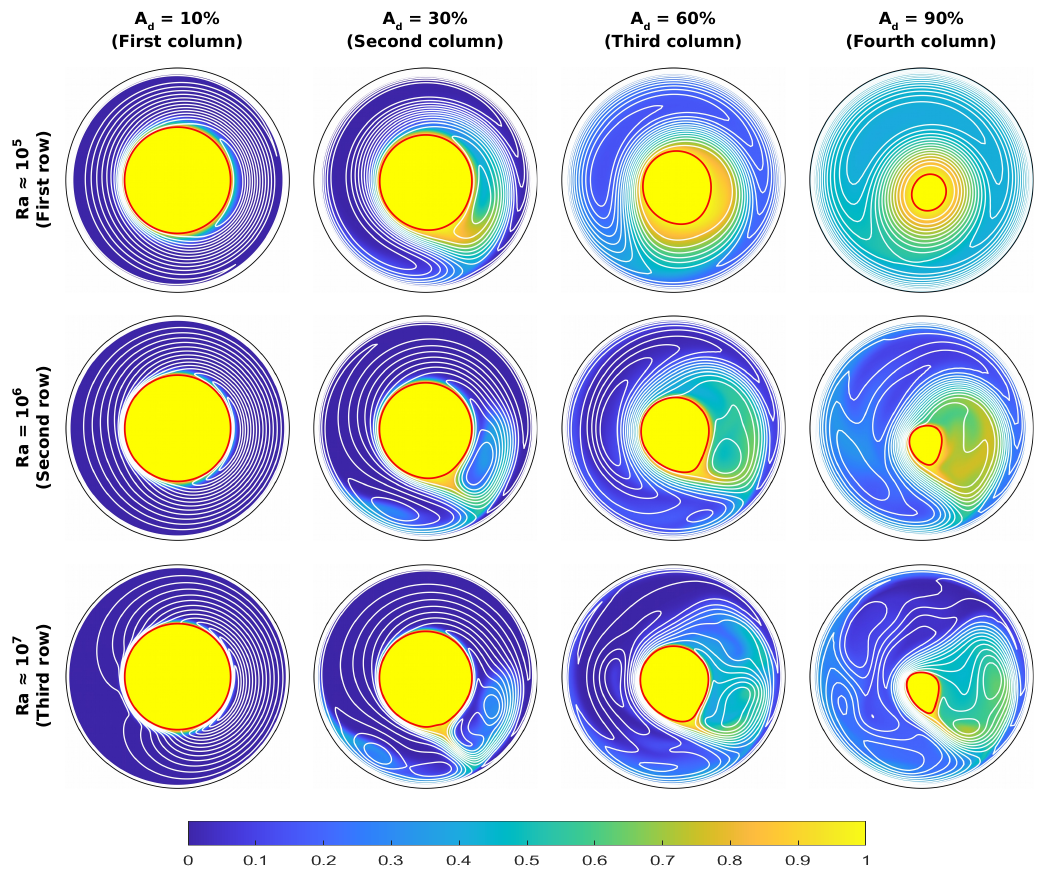}
  \caption{Streamlines (white contours) overlaying the concentration contours of the dissolved solute at $A_d(t) = 10\%$, $30\%$, $60\%$, and $90\%$ (left to right) for $\Omega =$1 and $Ra = 10^5$, $10^6$, and $10^7$ (top to bottom). The corresponding color map illustrates the concentration variations in the domain at these instances.} 
  \label{fig:flow_fixomg}
\end{center} 
\end{figure}

A similar analysis of the flow regimes is performed by varying the Rayleigh number ($Ra$) while keeping the rotation speed fixed at $\Omega = 1$, which corresponds to the case when the rotation-induced laminar velocity of the cylinder is the same as that of the buoyancy-induced characteristic velocity. Figure \ref{fig:flow_fixomg} depicts that the flow varies significantly as the Rayleigh number increases. For $Ra \geq 10^6$, flow tends to become more irregular, and dissolved solute is localized in the right half of the cylinder. 
 
\textcolor{blue}{
As discussed earlier, in the absence of rotation, buoyancy causes the dissolved solute-laden heavier fluid to sink near the solute boundary and lighter fluid to rise near the cylinder, respecting mass conservation. As dissolution increases, this convective flow progressively becomes confined within the lower half of the cylinder (See Supplemental Movie S1). 
When rotation is added to the cylinder, the upward flow near the cylinder is enhanced on the left. At the early stages of the dissolution, the qualitative feature of the flow remains similar on the left, barring fluid parcels moving faster near the cylinder, supported by the clockwise rotation of the cylinder. On the other hand, on the right, the upward flow of the fluid parcels near the cylinder is opposed by the clockwise rotating cylinder. As a result, a region of upward-moving fluid parcels is formed, confined between two regions with downward-moving fluid parcels. The former carries fluid parcels atop the undissolved solute, leading to a forced convective flow and more dissolution on the left. With an increasing dissolution of the solute, the clockwise flow (on the right) near the solute becomes progressively weaker, and the solute appears to be tilted on the right. We further noticed that the clockwise flow near the solute becomes weaker as the rotation speed increases, leading to only an anticlockwise flow near the solute and a clockwise flow near the cylinder. See Supplementary Movies S2-S4. 
}
Overall, we notice that rotation alters the flow topology that leads to the spreading of the dissolved solute around the interface instead of being localized only in the bottom half of the cylinder ($\Omega = 0$). This causes the concentration gradient near the upper part of the interface to be shallow, and hence, the dissolution slows down.


To further explore the behavior of the flow regimes, we trace trajectories of passive tracers released initially at the solute/fluid interface. These trajectories represent the respective pathlines of individual fluid particles coinciding with the tracer particle and reveal how the flow dynamics change accordingly. The equations governing the trajectory of a particle $P$ with position $(X_P(t), Y_P(t))$ are: 
\begin{equation}
    \label{eq:particle_ra5}
    \frac{dX_P}{dt} = u(X_P,Y_P,t),  \quad \frac{dY_P}{dt} = v(X_P,Y_P,t), 
\end{equation}
where the Eulerian velocity at its location gives the Lagrangian velocity of a particle. These equations are associated with initial conditions, $X_P(t = 0) = X_P^0, \; Y_P(t = 0) = Y_P^0$. 

\begin{figure}
 \begin{center}
 $\Omega = 0$ \hspace{2.3cm} $\Omega = 0.5$ \hspace{2.3cm} $\Omega = 1$ \hspace{2.3cm} $\Omega = 1.5$ \\  
 \includegraphics[width=\textwidth]{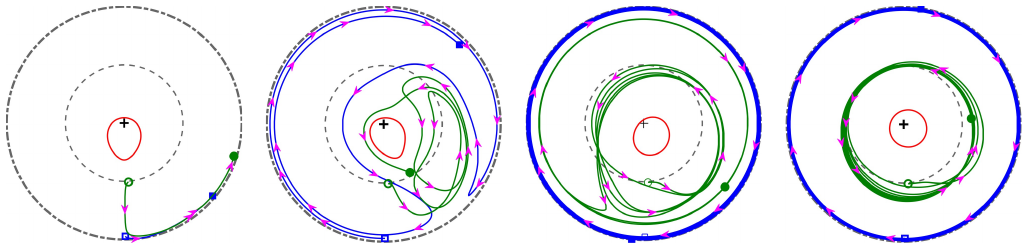}
 \end{center} 
\caption{Pathlines of two passive tracers, released initially at $(0.0745, -1.0115)$ and $(-0.0204, -1.9441)$, for rotation speeds ranging from $0$ to $1.5$ (shown left to right with an increment of $0.5$). The respective trajectories of these particles are shown in green and blue. The initial positions of the tracers are marked with empty markers, while their final positions are marked with the corresponding filled markers. 
Arrow markers along the pathlines denote the direction of particle motion. 
The red contour denotes the final shape of the undissolved solute, while $+$ represents the centre of the cylinder. 
} 
  \label{fig:particle}
\end{figure}

\textcolor{blue}{
Figure \ref{fig:particle} shows the trajectories of two tracer particles initially placed at $(X_P^0, Y_P^0) = (0.0745$, $-1.0115)$ and $(X_P^0, Y_P^0) = (-0.0204, -1.9441)$ for $Ra = 10^5$ and various rotation speeds ranging from 0 to 1.5. The initial positions of the tracers are chosen judiciously to capture the key features of the flow. The former (subsequently called as `P1') starts near the solute, where the flow variation is primarily dominated by buoyancy. On the other hand, the latter (subsequently called as `P2') is released near the cylinder, where the flow is dominated by rotation. The trajectories are tracked till 90\% dissolution of the solute is achieved. In the absence of rotation, P1 is driven by natural convection and follows a trajectory vertically downward before it moves upward along the cylinder wall, whereas P2 experiences upward motion only (see Supplementary Movie S5). 
}

\textcolor{blue}{
Cylinder rotation significantly alters the particle trajectories. Particles P1 and P2 primarily traverse in the anti-clockwise and clockwise directions, respectively. However, for each $\Omega$, we observe a unique and distinct characteristic of the particle trajectories. 
When buoyancy dominates rotation ($\Omega = 0.5$), the trajectory of P1 is confined to the right half of the cylinder for the majority of the dissolution, except at a later stage when the solute shrinks significantly and the particle has to traverse a small distance to end on the left half of the cylinder. Nonetheless, it eventually ends up on the right half (filled green marker) of the cylinder after spending a small duration on the left. For the same flow parameters, P2 initially rotates in the clockwise direction in accordance with the clockwise motion of the cylinder. However, before completing a full revolution around the solute, the particle is caught in the anti-clockwise rotating fluid parcels on the right, attracting P2 towards the solute. Subsequently, the particle traversed around the solute in the anti-clockwise direction and reenters the clockwise rotating boundary layer adjacent to the cylinder (see Supplementary Movie S6). 
}

\textcolor{blue}{
For $\Omega > 1$, when rotation dominates buoyancy, P1 remains confined within a neighbourhood of the initial interface position. In particular, for $\Omega = 1.5$, it is evident from figure \ref{fig:particle} that the particle exhibits a cyclic motion around the solute. The broken symmetry of the trajectory is attributed to the buoyancy that causes a departure of the particle from a perfect cyclic motion. However, as $\Omega$ increases, the asymmetry becomes less prominent, and the number of revolutions of the trajectory around the solute increases (not shown for brevity). See Supplementary Movie S7. 
}

\textcolor{blue}{
Finally, for $\Omega = 1$, we observe that initially, P1 follows a trajectory similar to $\Omega = 0.5$. However, instead of being confined within the right half of the cylinder, it revolves around the solute as time progresses. After a couple of revolutions around the solute, the particle is slowly pushed away from the interface by a strong buoyant force below the solute. This process continues for a while before the particle is eventually trapped within the boundary layer near the cylinder wall and traverses in the clockwise direction. 
As anticipated, for both $\Omega = 1$ and $\Omega = 1.5$, P2 rotates in the clockwise direction confined within the boundary layer near the cylinder. See Supplementary Movie S8. 
}

The above observations of particle paths indicate that the fluid adjacent to the cylinder wall flows in the clockwise direction, while the fluid adjacent to the interface experiences an overall anti-clockwise motion. For all the cases under consideration in figure \ref{fig:particle}, the pathlines are smooth, regular, and do not exhibit any disorderly behaviour, a signature of laminar flow. This behaviour is particularly evident at higher rotation speeds, for which the trajectories exhibit smooth circular motions around the solute. 

\subsection{Quantitative impacts of rotation and buoyancy on solute transport} \label{subsec:quantitative_solute}

In this section, we quantify dissolution, mixing, and interface shape due to nonlinear interactions between fluid flow and solute dissolution over a wide range of rotation speed ($\Omega$) and Rayleigh number ($Ra$). 


\subsubsection{Dissolution time of the solute}\label{subsubsec:dissolution}

Our numerical experiments are carried out until the solute experiences $95\%$ dissolution. Using these simulated data, we predict the time required for complete solute dissolution using spline extrapolation. Table \ref{table:total_time} lists the time for the complete dissolution for Rayleigh numbers varying by two orders of magnitude and $\Omega$ ranging from 0 to 1.5. 
Interestingly, rotation-induced forced convection does not enhance dissolution. Rather, for a fixed Rayleigh number, the dissolution process slows down as the rotation speed increases due to a shallow concentration gradient at the interface. 
We further note, for $Ra = 10^5$ and $10^6$, a rotation speed $\Omega = 1.5$ can significantly delay the complete dissolution of the solute with almost 85\%-94\% increase in complete dissolution as compared to $\Omega = 0$. Whereas, for $Ra = 10^7$, time taken for complete dissolution with $\Omega = 1.5$ increases only up to 8\%-9\% relative to the case of no rotation. In summary, the relative strength of rotation to buoyancy plays a critical role in determining the dissolution of the solute. It appears that there exists a critical rotation speed $\Omega_{cric}(Ra)$ such that for $\Omega < \Omega_{cric}(Ra)$, dissolution has a weak dependence on rotation. Our prediction further indicates that $\Omega_{cric}$ increases with $Ra$. 

\begin{table}
\begin{center}
\begin{tabular}{ lcccc } 
 \thead{Rayleigh number \\ ($Ra$)} &  \thead{Rotation speed \\ ($\Omega$)} &  \thead{$90\%$ dissolution time\\ (Numerical data)} &  \thead{$95\%$ dissolution time\\ (Numerical data)} &   \thead{$100\%$ dissolution time\\ (Predicted data)} \\
\multirow{4}{2em}{$10^5$} & 0   & 52.22  & 60.27  & 71.32  \\ 
                          & 0.5 & 52.25  & 59.52  & 69.37  \\ 
                          & 1.0 & 90.55  & 105.02 & 122.18 \\ 
                          & 1.5 & 105.40 & 120.15 & 137.93 \vspace{0.2 cm} \\
\multirow{4}{2em}{$10^6$} & 0   & 41.20 & 46.40 & 52.84  \\ 
                          & 0.5 & 42.12 & 47.30 & 53.33  \\ 
                          & 1.0 & 46.77 & 53.40 & 59.99 \\ 
                          & 1.5 & 66.75 & 82.37 & 97.91 \vspace{0.2 cm} \\
\multirow{4}{2em}{$10^7$} & 0   & 46.65 & 52.10 & 59.04  \\ 
                          & 0.5 & 47.85 & 53.45 & 60.11  \\ 
                          & 1.0 & 49.72 & 55.70 & 63.35 \\ 
                          & 1.5 & 50.50 & 56.30 & 63.90 \\ 
\end{tabular}
\caption{Time required for solute dissolution at various Rayleigh numbers and rotation speeds.}
\label{table:total_time}
\end{center}
\end{table}

\begin{figure}
  \centering
  \includegraphics[width=\textwidth]{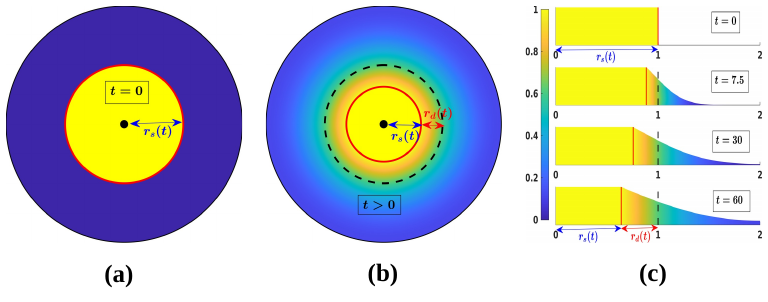}
  \caption{
    Concentration of dissolved and undissolved solute in the absence of rotation and buoyancy: (a) initially ($t = 0$) and (b) at a later time ($t > 0$). Dissolution recedes the interface (red) radially. Due to this rotational symmetry, the Stefan problem associated with the dissolution can be studied through a one-dimensional model. (c) The concentration profile in the radial direction is shown from top to bottom at $t = 0$, $t = 7.5$ (early time), $t = 30$ (intermediate time), and $t = 60$ (later time).
  }
  \label{fig:schematic_scaling}
\end{figure}

\begin{figure}
  \centering
  \begin{subfigure}[b]{0.32\textwidth}
    \includegraphics[width=\textwidth]{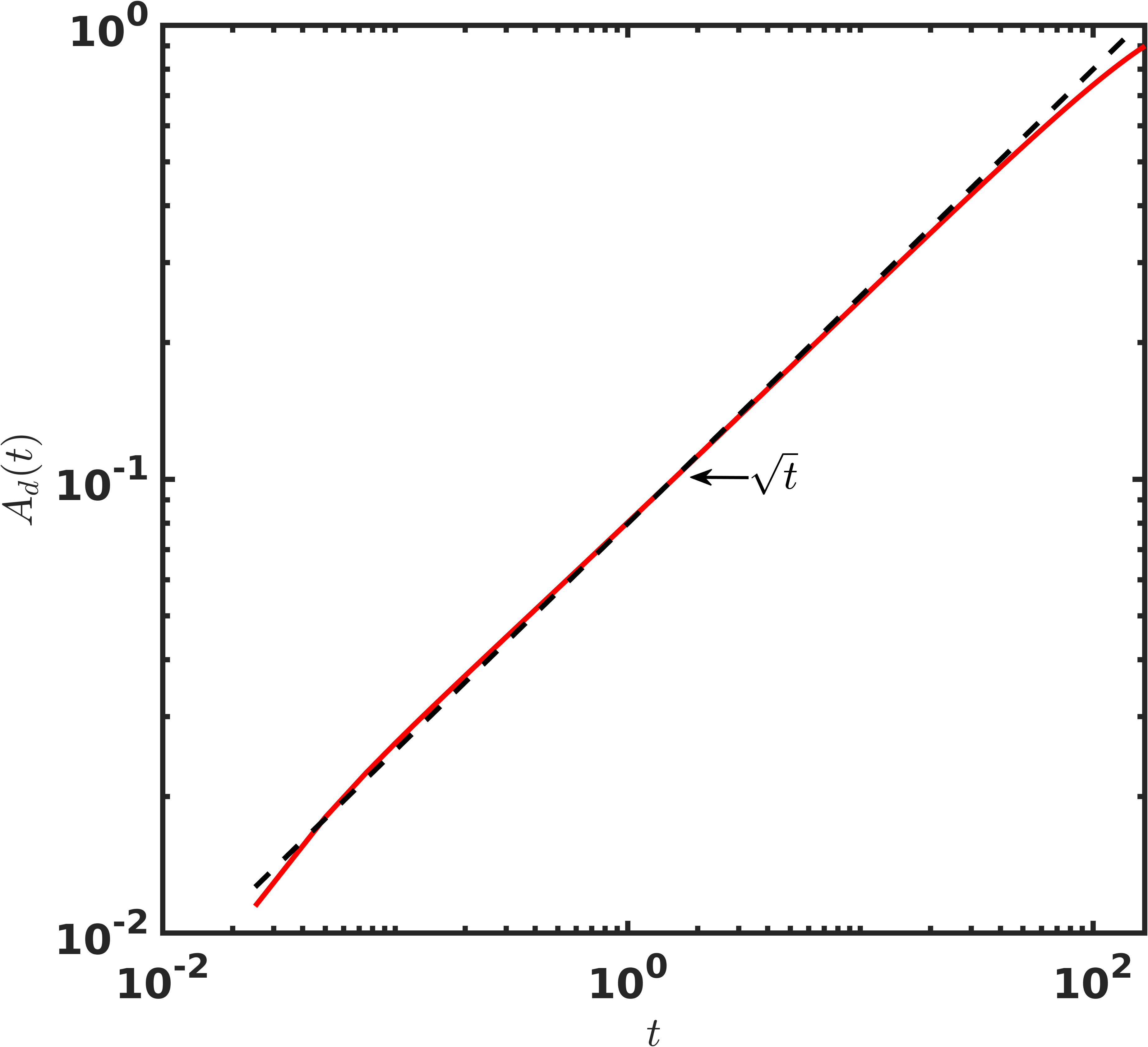}
    \caption{}\label{fig:nobnor}
  \end{subfigure}
  \hfill
  \begin{subfigure}[b]{0.32\textwidth}
    \includegraphics[width=\textwidth]{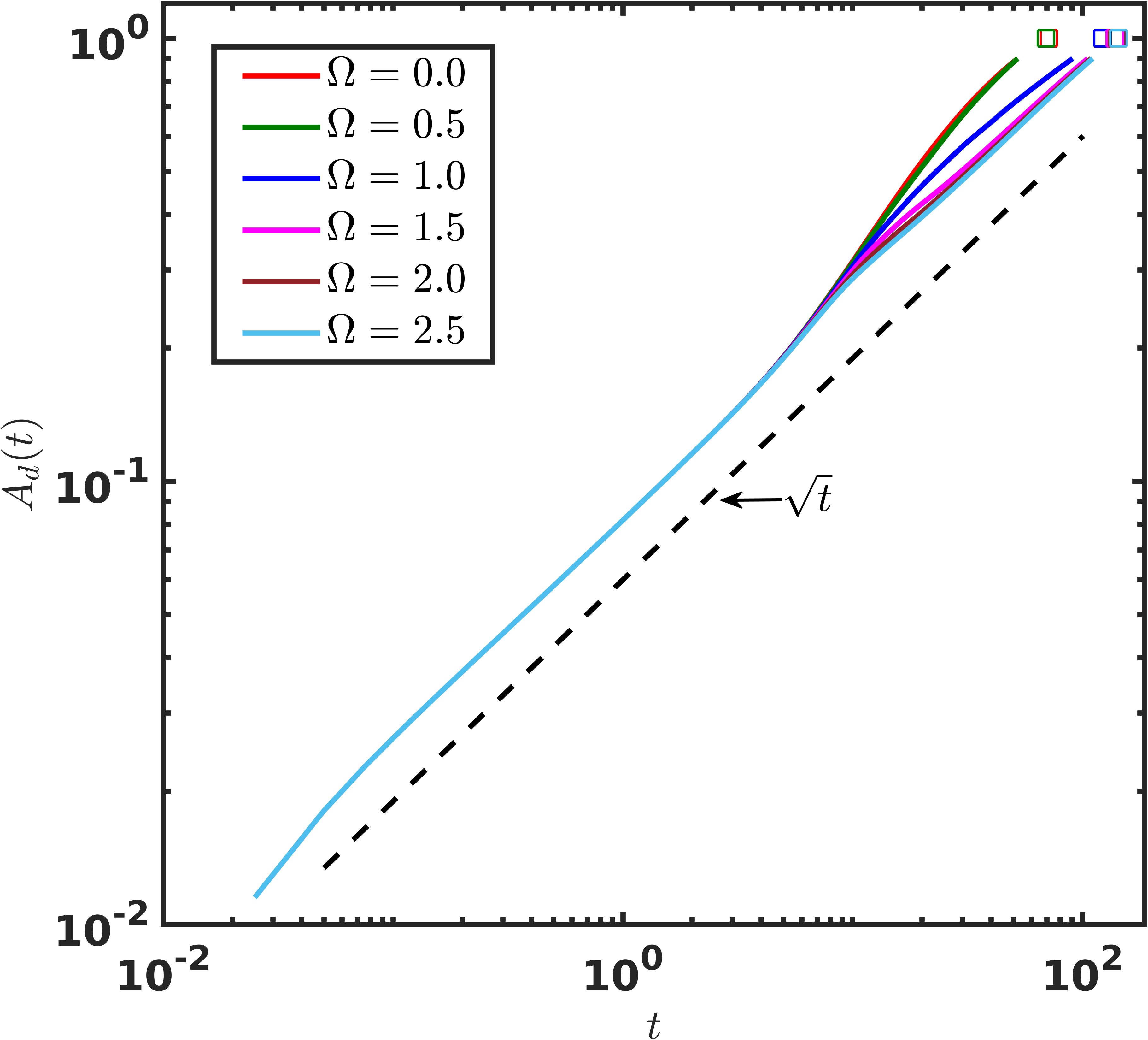}
    \caption{}\label{fig:fix_Ra}
  \end{subfigure}
  \hfill
  \begin{subfigure}[b]{0.32\textwidth}
    \includegraphics[width=\textwidth]{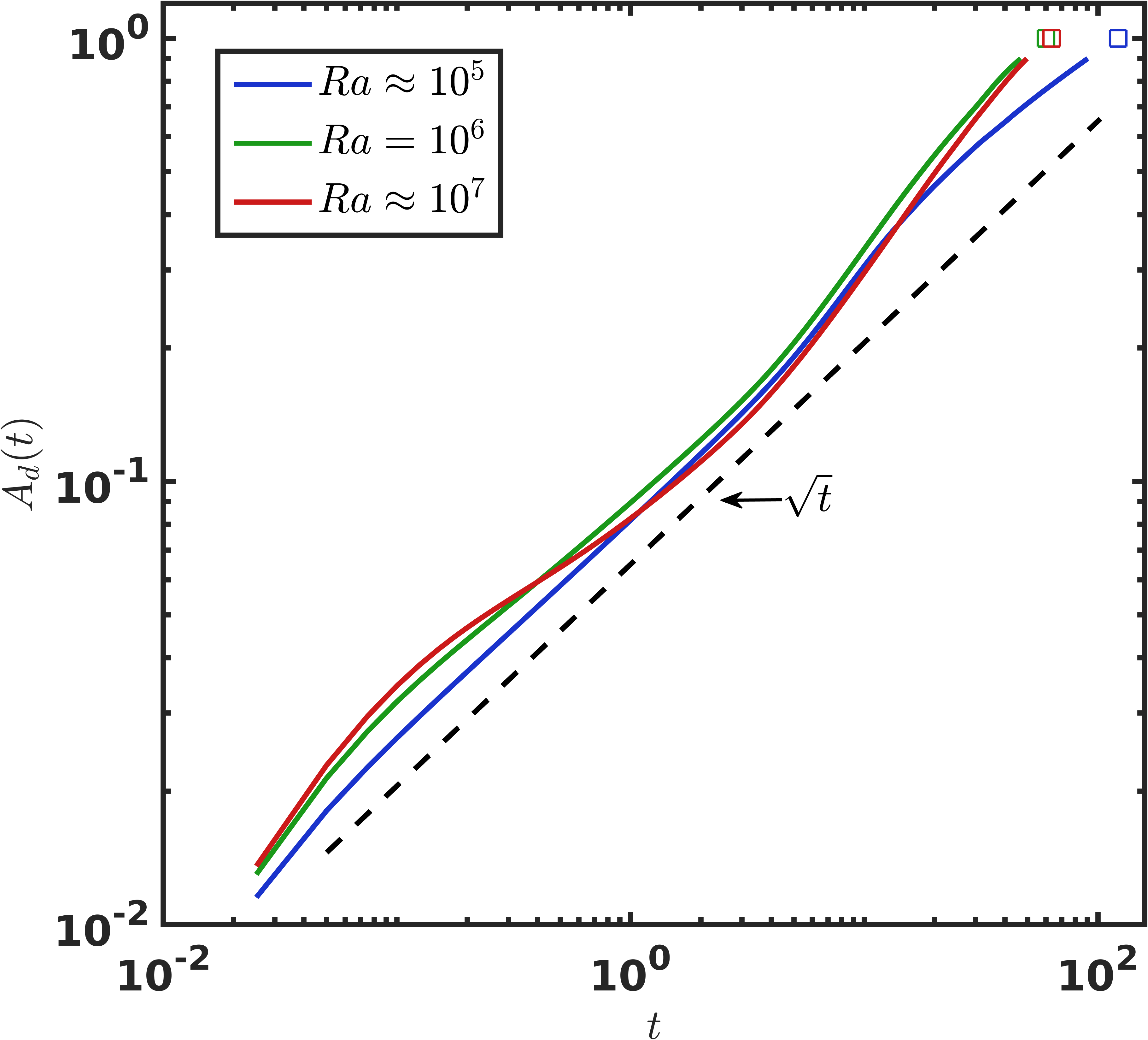}
    \caption{}\label{fig:fix_omg}
  \end{subfigure}
  \caption{Time required for the gradual dissolution of the solute: (a) without buoyancy or rotation, (b) for various rotation speeds at a fixed Rayleigh number $Ra \approx 10^5$, and (c) for various Rayleigh numbers at fixed rotation speed $\Omega = 1$. Solid curves represent numerical results up to 90\% dissolution. The dashed line, proportional to $\sqrt{t}$, is shown as a guide to the eye. Square markers denote the predicted time for complete dissolution given in Table \ref{table:total_time}.}
  \label{fig:dissolution_time}
\end{figure}

We now aim to theoretically estimate a relationship for the solute dissolution time. To achieve this, we first consider a scenario where the effects of buoyancy and rotational forces are neglected. In such cases, the chosen problem is driven solely by diffusion forces, which reduces it to a classical one-dimensional phase-change problem in the radial direction due to the symmetric nature of the system. 
Let $r_s(t)$ and $r_d(t)$ denote the radius of the undissolved solute and the distance traversed by the solid-liquid interface from its initial position, respectively, at an instantaneous time $t$ (see figure \ref{fig:schematic_scaling}). Thus, we have
\begin{equation}
    \label{eq:radius}
    r_d(t) + r_s(t) = 1 \quad \text{with} \quad r_d(0) = 0 \quad \text{and} \quad r_s(0) = 1. 
\end{equation}
Overall, the problem can be viewed as the dissolution of a solute block in a fluid, with an expanding fluid region. Resorting to $c(r = r_d(t), t) = 1$ and following the analysis of \citet{stefan1891theorie}, we obtain $r_d(t)\propto \sqrt{t}$. Thus, from \eqref{eq:radius} we derive, 
\begin{equation}
    \label{eq:dAtime}
    1-r_s(t) = k \sqrt{t} \quad \mbox{which yields} \quad A_d(t) = 2k\sqrt{t} - k^2t, 
\end{equation}
where $k$ is a constant of proportionality. The above-mentioned scaling relation is valid until the dissolved solute reaches the cylinder wall. Therefore, in the present study, the finite-size effect of the fluid region leads to a breakdown of the said power-law behavior. We performed numerical simulations by neglecting buoyancy and rotation in our model formulation and determined the interface evolution, which agrees excellently with $k \sqrt{t}$ for $k = 0.08$ (correct up to 2 decimal places) up to $t \lesssim 30$ (see figure \ref{fig:nobnor}).

Effects of $Ra$ and $\Omega$ on the evolution of $r_d(t)$ and hence $A_d(t)$ are investigated next. We numerically compute $A_d(t)$, defined in equation \eqref{eq:Ad}, for different values of $Ra$ and $\Omega$. For a fixed $Ra$, the evolution of $A_d(t)$ is independent of $\Omega$ at the early stages of the dissolution. For example, with $Ra = 10^5$, the effects of rotation become evident at $A_d(t) \gtrsim 0.25$ (see figure \ref{fig:fix_Ra}). Provided the evolution of the dissolved solute area satisfies the relation \eqref{eq:dAtime}, one infers that $k$ depends on $\Omega$ in such a way that with increasing $\Omega$, $k^2 t$ is dominated by $2 k \sqrt{t}$ in \eqref{eq:dAtime}. This is attributed to the fact that at higher rotation speeds, the rotation-induced effects dominate, and the buoyancy effect becomes negligible, causing the system to behave more like a purely diffusion-driven problem. On the other hand, for a fixed $\Omega$, the effects of $Ra$ are evident throughout the dissolution process. However, the overall evolution of $A_d(t)$ can be described by a power-law relation $A_d(t) \sim t^a$, for $a \in (0.5278, 0.6354)$.

In an attempt to explain this nonlinear dependence on $Ra$ and $\Omega$, we revisit the non-dimensional parameter $Ra$ and $\Omega$. In a forced convection, the velocity field is proportional to $\Omega$. Thus, we write
\begin{equation}
    \label{eq:Ra_Omg_relation}
    Re \sim \Omega, \quad Pe \sim \Omega \quad \mbox{resulting} \quad Ra \sim \Omega^2.
\end{equation}
Subsequently, we observe that $A_d(t)$ follows a $\sqrt{t}$ relation provided 
\begin{equation}
    \label{eq:Ra_Omg}
    \sqrt{Ra_{\Omega}} \lesssim 250  \quad \mbox{ where} \quad Ra_\Omega = \frac{Ra}{\Omega^2}. 
\end{equation}
Within this specific range for $Ra_\Omega$, the flow is dominated by rotation, and the interface is consistently observed to maintain its nearly initial circular shape throughout the dissolution process. On the other hand, when $\sqrt{Ra_{\Omega}} > 250$, the interface loses its circular shape due to a dominant buoyancy force (see 2$^{\rm nd}$ and 3$^{\rm rd}$ rows of figure \ref{fig:flow_fixRa} and 2$^{\rm nd}$ and 3$^{\rm rd}$ panels of figure \ref{fig:solute_shape_Omega}).

\subsubsection{Mixing of the dissolved solute} \label{subsubsec:mixing} 

Here, we quantify mixing of dissolved solute in the solvent and explore the roles of buoyancy and rotation, varying $Ra$ and $\Omega$. The degree of mixing is defined as \citep{BirendraJha2011} 
\begin{equation}
    \label{eq:degree_mixing}
    \chi(t) = 1 - \frac{\sigma^2(t)}{\sigma^2_{max}},
\end{equation}
where the variance 
\begin{equation}
    \label{eq:variance}
    \sigma^2(t) = \langle c^2 \rangle - {\langle c \rangle}^2,
\end{equation}
of the concentration field can be calculated in terms of the spatial average defined as
\begin{equation}
    \label{eq:average_c}
    \langle \ast \rangle (t) = \int_{\Sigma_l(t)} \ast (x,y,t) \, dx \, dy. 
\end{equation}

Here, $\sigma^2_{max}$ represents the maximum concentration variance observed during the dissolution or mixing process. Equation \eqref{eq:degree_mixing} signifies that at the perfectly segregated state ($\chi = 0$) variance attains its maximum value; whereas, the perfectly mixed state, at which concentration is uniform throughout the fluid region, ($\chi = 1$) corresponds to the zero variance.

\begin{figure}
  \centering
  \begin{subfigure}[b]{0.48\textwidth}
    \includegraphics[width=\textwidth]{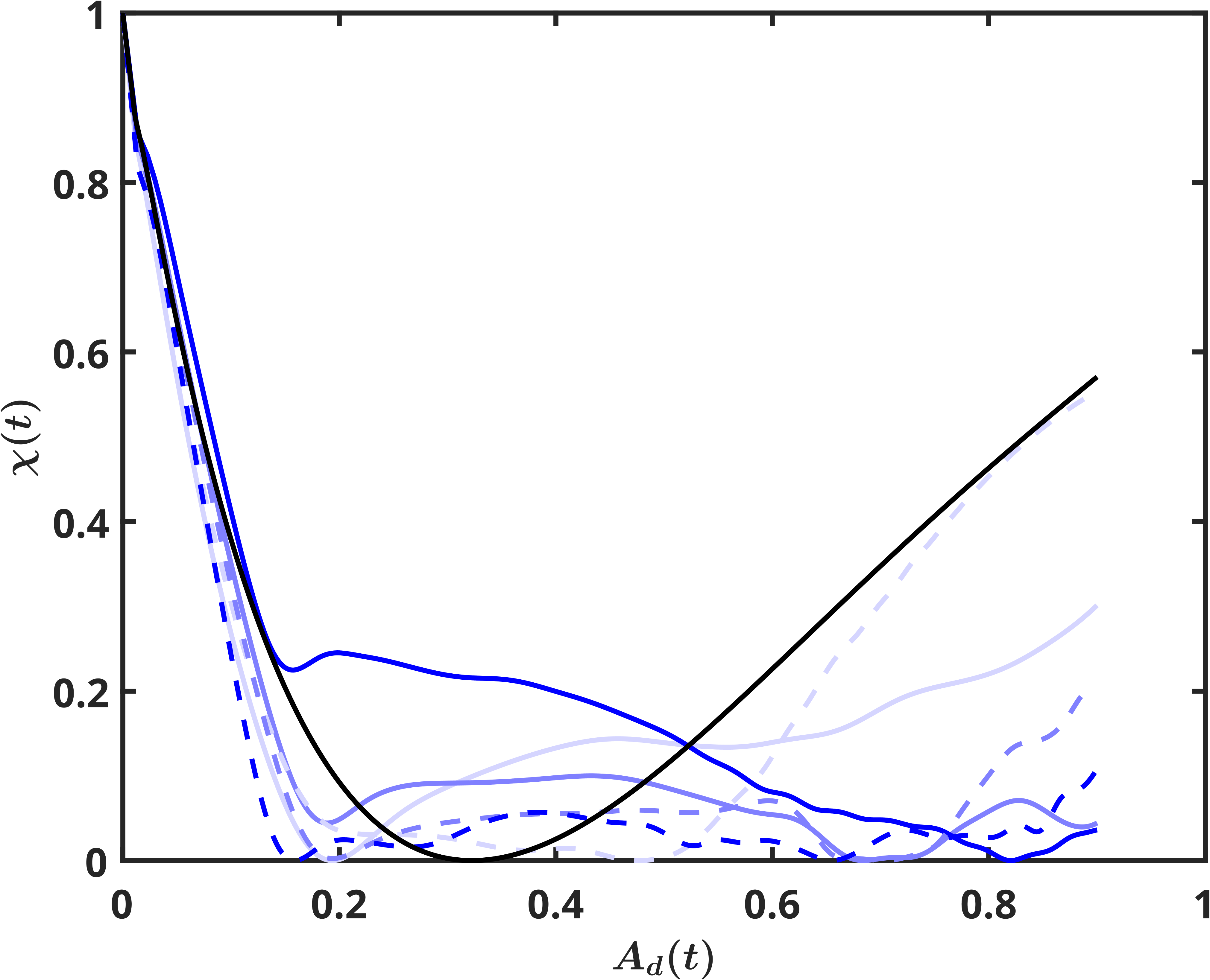}
    \caption{}\label{fig:mixing_fixRa}
  \end{subfigure}
  \hfill
  \begin{subfigure}[b]{0.48\textwidth}
    \includegraphics[width=\textwidth]{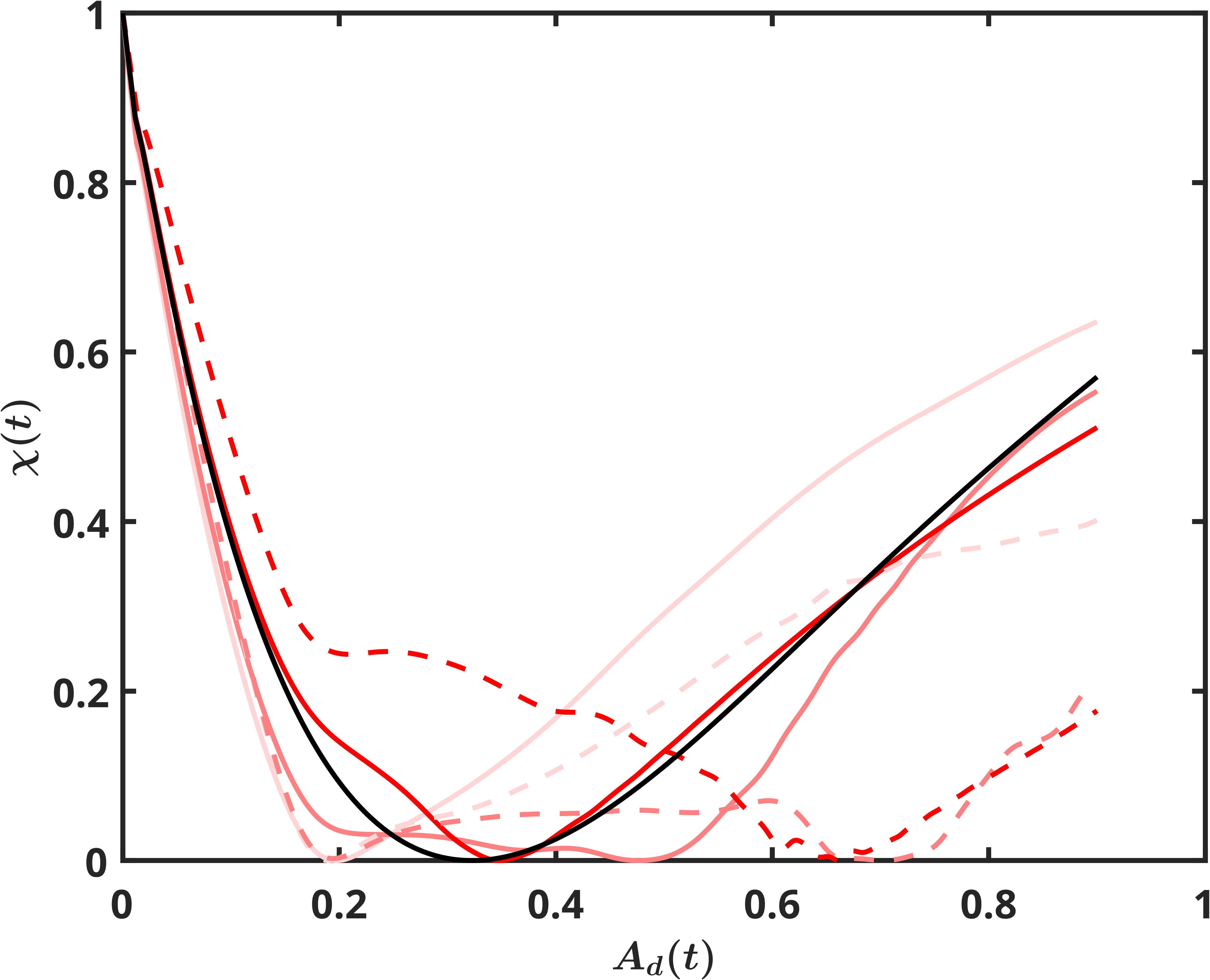}
    \caption{}\label{fig:mixing_fixOmg}
  \end{subfigure}
  \caption{
    Effects of buoyancy and rotation on the variation of the degree of mixing with the amount of dissolved solute, $A_d$: (a) $Ra = 10^5$, $10^6$, and $10^7$ (light to dark), without rotation ($\Omega = 0$, solid line) and with rotation ($\Omega = 1$, dashed line); (b) $\Omega = 0.5$, $1$, and $2$ (light to dark) for $Ra = 10^5$ (solid line) and $Ra = 10^6$ (dashed line). The black line corresponds to pure diffusion, i.e., without buoyancy or rotation.
  }
  \label{fig:mixing}
\end{figure}

Due to the absence of dissolved solute in the fluid region initially ($c_0(\vec{x}, 0) \equiv 0$), spatial variance is zero and hence $\chi(0) = 1$. As solute dissolves and it mixes with the solvent, the $\sigma^2(t)$ increases, resulting in $\chi(t)$ to decrease until it reaches a local minimum, $\chi_{min}^{local, 1}$. We denote the corresponding dissolved solute area using $A_{d, min}^{local, 1}$. For $A_d(t) \leq A_{d, min}^{local, 1}$, $\chi(t)$ is dominated by diffusion with a decay rate that depends on $Ra$ and $\Omega$. We note that, depending on $Ra$ and $\Omega$, $\chi(t)$ may attain the global minimum value $\chi_{min}^{global} (= 0)$. The smallest $A_d(t)$ for which $\chi(t) = 0$ is denoted as $A_{d, min}^{global}$. For the reference case of no rotation and buoyancy (black line in figure \ref{fig:mixing}), $A_{d, min}^{local, 1} = A_{d, min}^{global}$, which may also occur for certain combinations of $Ra$ and $\Omega$. In the absence of rotation ($\Omega = 0$), $A_{d, min}^{global}$ increases as $Ra$ increases (solid lines in figure \ref{fig:mixing_fixRa}). On the other hand, for $\Omega = 1$, $A_{d, min}^{global}$ decreases as $Ra$ increases (dashed lines in figure \ref{fig:mixing_fixRa}). However, no such correlation is obtained varying $\Omega (>0)$ for a fixed $Ra$ (see figure \ref{fig:mixing_fixOmg}). Evolution of $\chi(t)$ beyond $A_{d, min}^{local, 1}$ is governed by nonlinear interactions of buoyancy and rotation forces. Nonetheless, for a given $Ra$, at 90$\%$ dissolution of the solute, a better mixed state is obtained through rotation as compared to that in the absence of rotation (figure \ref{fig:mixing_fixRa}). Likewise, for a given rotation speed $\Omega$, at 90$\%$ dissolution of the solute, a better mixed state is obtained for smaller $Ra$ (figure \ref{fig:mixing_fixOmg}). Thus, we conclude that there is an optimal pair of $Ra$ and $\Omega$ that causes the dissolved solute to mix better in the solvent. 

\subsubsection{Symmetry breaking due to rotation}\label{subsubsec:asymmetry} 

As discussed in \S \ref{subsec:qualitative_flow_dissolution}, in the absence of rotation, the dissolved solute shrinks due to buoyancy force, spreads symmetrically about the $y$-axis, and primarily occupies the bottom half of the cylinder -- center of mass of the dissolved solute is (0, $-y^\ast(t; Ra)$). The introduction of the cylinder rotation breaks this symmetry. To investigate this symmetry breaking, we compute the center of mass of the dissolved solute as follows: 
\begin{equation}
    \label{eq:meanpoint}
    \mu_X(t) = \int_{\Sigma_l(t)} x \, c_X(x,t) \, dx, \quad \mu_Y(t) = \int_{\Sigma_l(t)} y \, c_Y(y,t) \, dy, 
\end{equation}
where 
\begin{eqnarray}
    & \displaystyle c_X(x, t) = \int_{\Sigma_l(t)} c_{X,Y}(x, y,t) \, dy, \label{eq:mgpdf-x} \\ 
    & \displaystyle c_Y(y, t) = \int_{\Sigma_l(t)} c_{X,Y}(x, y,t) \, dx, \label{eq:mgpdf-y} \\ 
    \mbox{and} & \nonumber \\ 
    & \displaystyle c_{X,Y}(x, y,t) = \frac{c(x, y,t)}{\langle c \rangle (t)}. \label{eq:jpdf}
\end{eqnarray}

\begin{figure}
  \centering
  \begin{subfigure}[b]{0.48\textwidth}
    \includegraphics[width=\textwidth]{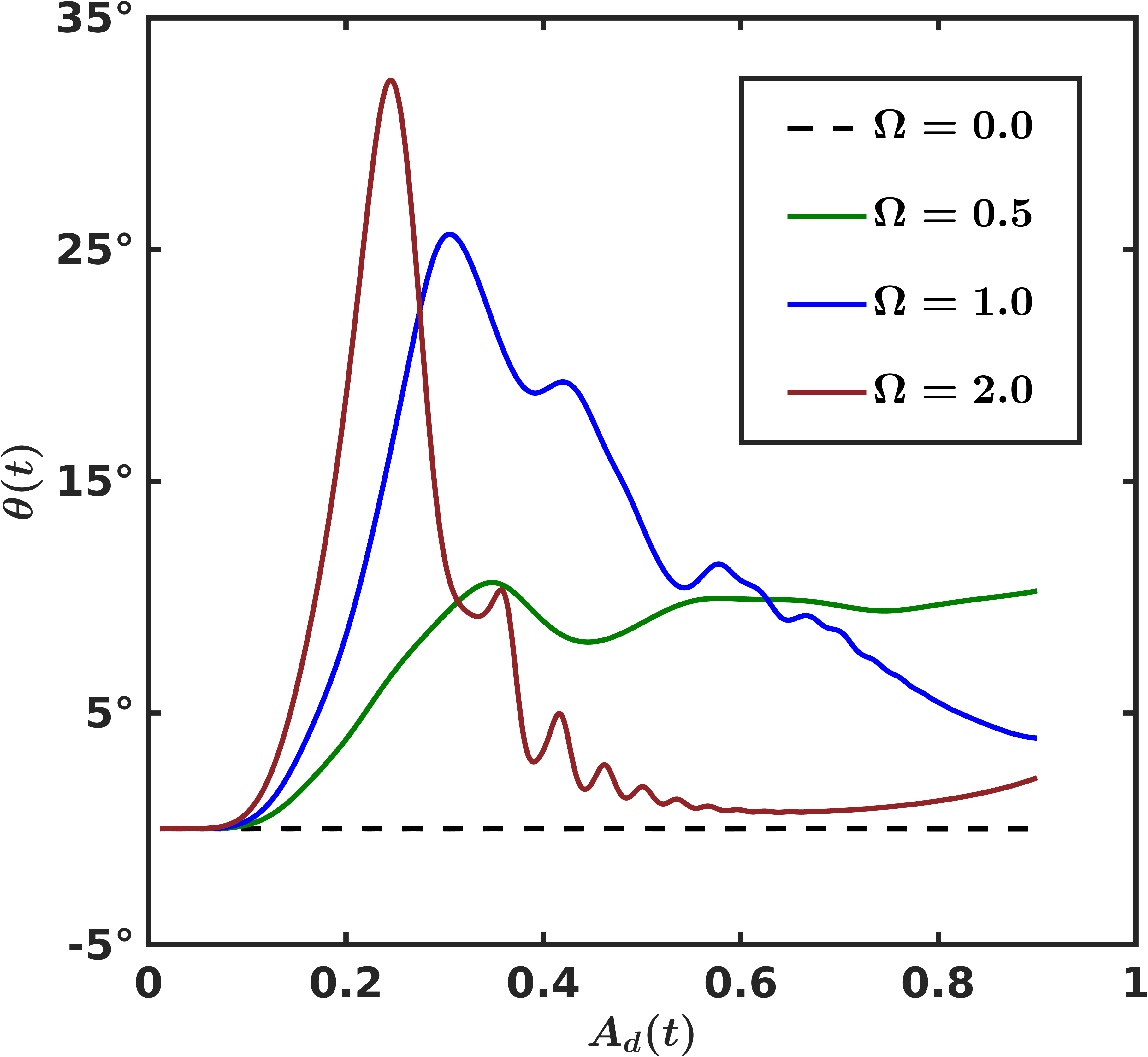}
    \caption{}\label{fig:angleRa}
  \end{subfigure}
  \hfill
  \begin{subfigure}[b]{0.48\textwidth}
    \includegraphics[width=\textwidth]{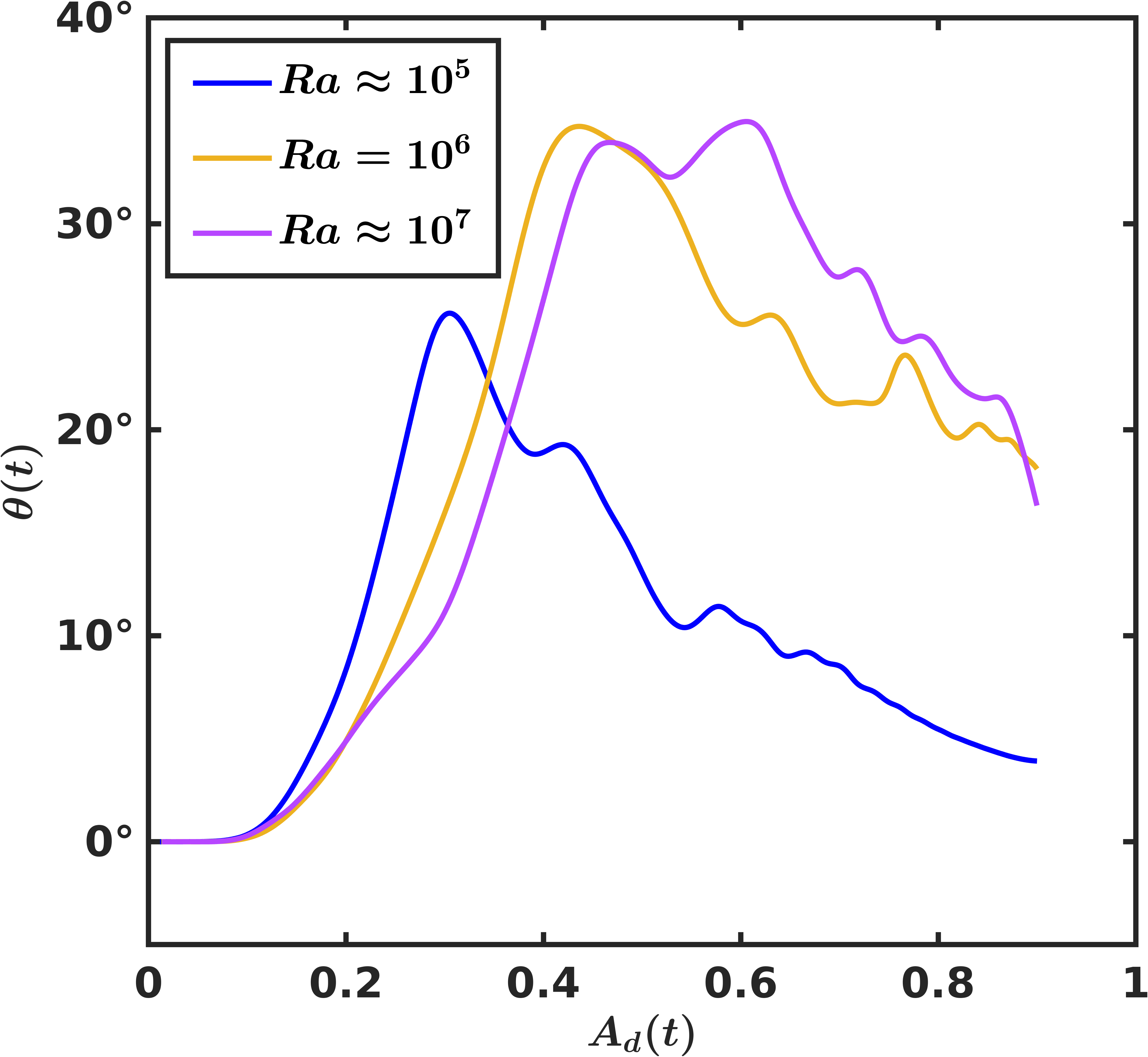}
    \caption{}\label{fig:angleOmg}
  \end{subfigure}
  \caption{
    Variation of the angle of inclination, $\theta(t)$, with the dissolved volume, $A_d(t)$: (a) fixed Rayleigh number $Ra = 10^5$ and varying rotation speed $\Omega$, and (b) fixed $\Omega = 1$ with varying $Ra$.
  }
  \label{fig:angle}
\end{figure}

The angular departure of the line joining ($\mu_X(t), \mu_Y(t)$) to the center of the cylinder from the negative $y$-axis, measured in the counterclockwise direction, can be computed as
\begin{equation}
    \label{eq:angle}
    \theta(t) = \tan^{-1}\left(\frac{\mu_Y(t)}{\mu_X(t)}\right) - \frac{3\pi}{2}. 
\end{equation}
As anticipated, in the absence of rotation, the inclination angle is always zero (see dashed line in figure \ref{fig:angleRa}). We discuss our observations for $Ra \in \{ 10^5, 10^6, 10^7 \}$, and $\Omega \in \{ 0.5, 1, 2 \}$. For all the parameter values considered here, $\theta$ remains non-negative, signifying that the center of mass of dissolved solute always shifts to the opposite direction of rotation of the cylinder. Though the overall variation of $\theta(t)$ with $A_d(t)$ is non-monotonic, at the early stages of dissolution, growing angular departure is dominated by diffusion and rotation. As the amount of dissolved solute increases in the solvent, buoyancy forces become stronger and counter the rotation. Nonlinear interactions of these two counterbalancing forces on the spreading of the dissolved solute result in non-monotonic variations of $\theta(t)$ with $A_d(t)$. Each local optimum value $\theta_{opt}^{local, i}, \; (i = 1, 2, \hdots)$ corresponding to a specific dissolution volume $A_{d, opt}^{local, i}$ signifies a reversal of dominance between rotation and buoyancy. To be more specific, for $A_{d, opt}^{local, 1} < A_d(t) < A_{d, opt}^{local, 2}$ buoyancy dominates rotation and $\theta(t)$ decreases, whereas, for $A_{d, opt}^{local, 2} < A_d(t) < A_{d, opt}^{local, 3}$, rotation dominates buoyancy and $\theta{(t)}$ increases and so on.

For $Ra = 10^5$, $A_{d, opt}^{local, 1}$ ($\theta_{opt}^{local, 1}$) decreases (increases) with $\Omega$ (see figure 
\ref{fig:angleRa}). On the other hand, for $\Omega = 1$, both $A_{d, opt}^{local, 1}$ and $\theta_{opt}^{local, 1}$ increase with $\Omega$ (see figure \ref{fig:angleOmg}). We further note that for $Ra = 10^5$ and $\Omega = 0.5$, $\theta(t)$ does not vary significantly 
as solute dissolves. As rotation increases $(\Omega \geq 1)$, competing buoyancy and rotation forces result in a rapid oscillation in $\theta$ as solute dissolves, and the frequency of oscillations increases with $\Omega$ (see figure \ref{fig:angleRa}). At a higher rotation speed, $\theta(t)$ decreases as $A_d(t)$ increases, signifying a homogeneous distribution of the dissolved solute in the fluid region, which is in agreement with our observations in \S \ref{subsec:qualitative_flow_dissolution}.

\subsubsection{Interface evolution} \label{subsubsec:interface_shape}

\begin{figure}
 \begin{center}
 $\Omega = 0$ \hspace{1.6 cm} $\Omega = 0.5$ \hspace{1.6 cm} $\Omega = 1$ \hspace{1.6 cm} $\Omega = 1.5$ \hspace{1.6 cm} $\Omega = 2$ \\ 
 \includegraphics[width=\textwidth]{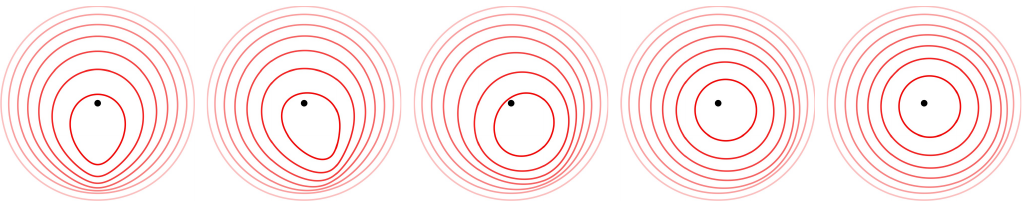}
 \end{center}
  \caption{The shape dynamics of the interface for rotation speeds ranging from $0$ to $2$ (shown from left to right with an increment of $0.5$) and $Ra = 10^5$. For each case, the positions of the interface evolutions are shown (inward with color changing from light to dark) at every 15\% interval of dissolution with the outermost curve representing the initial interface and the innermost at 90\% dissolution. The bullet marker indicates the center of the cylinder.} 
  \label{fig:solute_shape_Omega}
\end{figure}

This section examines how the shape of the interface evolves during the dissolution process at various rotation speeds. Figure \ref{fig:solute_shape_Omega} shows the interface at 15\% dissolution intervals up to 90\% for each rotation speed ranging from $0$ to $2$. Irrespective of the rotation speed, the interface at the initial stages (up to 15\% dissolution) maintains its circular shape and attains different shapes at later stages depending on the resultant of the forces.

In the absence of rotation, the interface evolves to an egg-like shape as solute dissolves. For the same parameter values, a qualitatively similar evolution has been observed by \citet{HUANG}, in which a circular solute placed at the center of a rectangular domain developed an egg-like shape during dissolution. 
At these stages, it appears that the solute moves downward due to buoyancy. However, in reality, the solute does not move from its initial position; rather, the convection-induced preferential dissolution of the upper part of the solute results in an apparent movement of the solute. A similar phenomenon is observed at a lower rotation speed of the cylinder ($\Omega \leq 0.5$). In such cases, the solute is seen to be tilted noticeably in the direction opposite to that of rotation. However, as rotation speed increases ($\Omega \geq 1$), the interface regains its circular shape. When the cylinder rotates sufficiently fast (e.g., $Ra = 10^5$, $\Omega \geq 1.5$), the interface maintains almost a perfectly circular shape exhibiting the axisymmetry of the flow and dissolution. A closer look reveals a slight asymmetry of the interface, which is in agreement with a non-zero $\theta$ for such cases. 

\begin{figure}
 \begin{center}
  $Ra = 10^5$ \hspace{3.3 cm} $Ra = 10^6$ \hspace{3.3 cm} $Ra = 10^7$ \\ 
 \includegraphics[width=\textwidth]{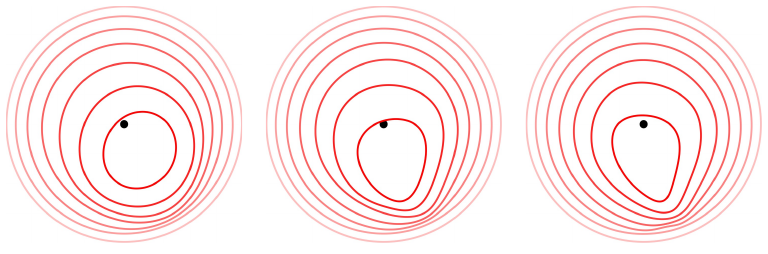}
 \end{center}
  \caption{The shape dynamics of the interface for $Ra = 10^5$, $10^6$ and $10^7$ (left to right) at a fixed rotation speed $\Omega = 1.0$. For each case, the positions of the interface evolutions are shown (inward with color changing from light to dark) at every 15\% interval of dissolution, with the outermost curve representing the initial interface and the innermost at 90\% dissolution. The bullet marker indicates the center of the cylinder.} 
  \label{fig:solute_shape_Ra}
\end{figure}

A similar analysis is of interface evolution for $\Omega = 1$ and varying $Ra \in \{ 10^5, 10^6, 10^7 \}$ indicates the axisymmetry of the interface breaks down as the strength of the buoyancy forces ($Ra$) increases (see figure \ref{fig:solute_shape_Ra}). As discussed in \S \ref{subsubsec:dissolution}, dissolution of the solute exhibits identical power-law behavior for $R_\Omega \lesssim 250$. Motivated by this observation, we investigated the evolution of the interface in terms of $Ra_\Omega$. As discussed earlier, we have seen through several experiments that the interface always exhibits a circular shape when $\sqrt{Ra_{\Omega}} \lesssim 250$ tends to lose this shape for higher values of $Ra_{\Omega}$. Here, we summarize the results of nine different numerical experiments in terms of three different values of the non-dimensional group $Ra_\Omega > 250$.  

\begin{figure}
 \begin{center}
 \includegraphics[width=\textwidth]{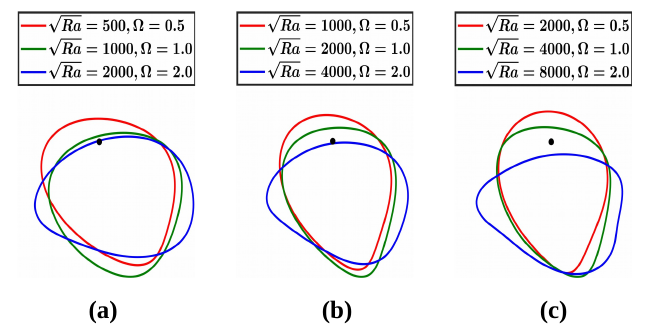}
  \caption{Shape dynamics of the interface at 90\% dissolution for three different modified Rayleigh numbers, (a) $\sqrt{Ra_{\Omega}} = 1000 $, (b) $\sqrt{Ra_{\Omega}} = 2000 $ and (c) $\sqrt{Ra_{\Omega}} = 4000 $. The bullet marker indicates the center of the cylinder.}
  \label{fig:solute_shape_RaOmega}
\end{center}
\end{figure}

Figure \ref{fig:solute_shape_RaOmega} shows an egg-like interface for all nine cases considered here. However, the orientation of the nose (the pointy end of the egg-like shape) of the interface is primarily determined by $\Omega$. As expected, the same rotation speed results in a similar interface and a similar orientation of the nose of the interface. Additionally, as $Ra_{\Omega}$ increases, the nose of the interface becomes sharper, exhibiting behavior similar to the trend observed when $Ra$ increases, as discussed in Figure \ref{fig:solute_shape_Ra}. It is worth mentioning that the above-mentioned behavior of the interface is limited to $Ra < 10^8$, for which the overall dynamics of the flow field and the solute dissolution exhibit significantly different features, and they are briefly discussed in the next section. 

\subsection{Dissolution dynamics at Ra = $10^8$}

\begin{figure}
 \begin{center}
 \includegraphics[width=\textwidth]{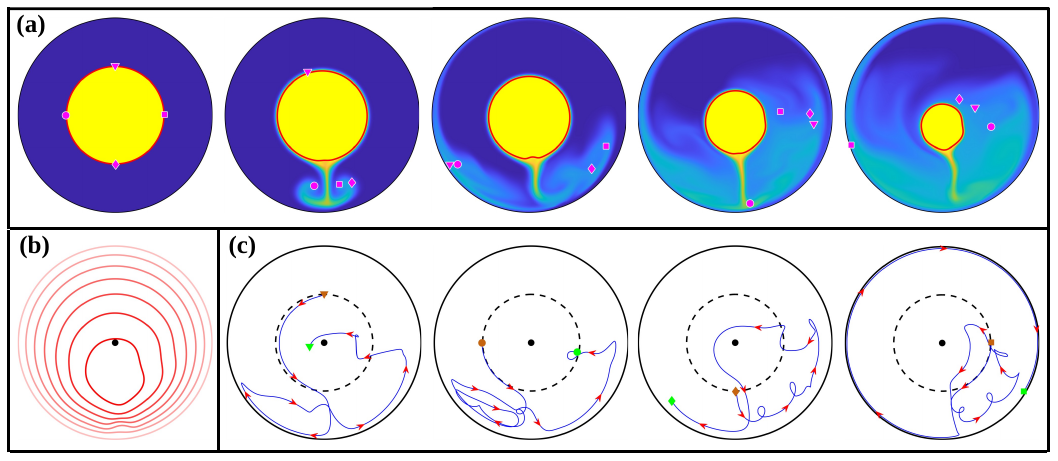}
  \caption{(a) Concentration (undissolved and dissolved solute) distribution at $0\%$ (initial), $15\%$, $30\%$, $60\%$, and $80\%$ dissolution (from left to right) for $Ra=10^8$ and $\Omega=0.5$. In each panel, the current position of four particles released at four different positions (left panel) is shown. (b) Shape of the interface at different dissolution volumes from $0\%$ to $90\%$ with an increment of $15\%$ (inward with color changing from light to dark). (c) Pathlines of passive tracers released at four different positions. In each case, the brown marker represents the starting position, while the green marker indicates the position at $90\%$ dissolution. 
  }
 \label{fig:ra8fig}
\end{center} 
\end{figure}

Throughout this paper, we have discussed results for the parameter range $10^5 \leq Ra \leq 6.4 \times 10^7$ and $0 \leq \Omega \leq 2.5$, for which the flow remains regular and exhibits a laminar solutal boundary layer. We close this section by summarizing numerical simulations for $Ra = 10^8$ and $\Omega = 0.5$ in figure \ref{fig:ra8fig}. We observed that the interface adopts an uneven shape such that the solute region is concave. The solutal boundary layer exhibits characteristics inconsistent with laminar boundary layers (see figure \ref{fig:ra8fig}a and figure \ref{fig:ra8fig}b). To better understand this phenomenon, we followed the trajectories of four tracer particles -- placed initially on the interface at angular locations 0, $\pi/2$, $\pi$ and $3 \pi/2$ -- as the solute dissolves. The trajectories of these tracers are captured up to 90\% of the solute dissolution and shown in figure \ref{fig:ra8fig}c. The loopy and relatively non-smooth particle trajectories indicate an irregular flow. Further quantification of these irregular flow dynamics is beyond the scope of the current study and will be considered in our future research. 

\section{Summary and discussions} \label{sec:discussion}
 
This work is concerned with the dissolution of a rod-shaped solute in a horizontal cylinder rotating in the clockwise direction. For low solutal Rayleigh numbers, the flow remains regular, resulting in a laminar solutal boundary layer regardless of the rotation speed. 
However, increasing the rotation speed renders the flow free of vorticity variation. 
An analysis of the dissolution time reveals that the amount of solute dissolved over time ($A_d(t)$) follows a power-law relation, which is determined by the balance between $2 k \sqrt{t}$ and $k^2 t$ terms. In the absence of rotation and buoyancy, a least-square fit to the numerical data yields $k \sim \mathbf{O}(10^{-1})$, confirming a $\sqrt{t}$ evolution of $A_d(t)$. Buoyancy and rotational convection exhibit a nonlinear influence on the amount of solute dissolved. For a fixed Rayleigh number, the square-root behaviour is maintained at higher rotation speeds. On the other hand, for a fixed rotation speed, the exponent of the power-law of the relation ranges between 0.5378 to 0.6354 depending on the Rayleigh number. 

During the dissolution process, as the rotation speed increases, the center of mass of the dissolved solute moves upward, countered by the buoyancy forces. For sufficiently large rotation speed, the dissolved solute is spread uniformly within the fluid region, bringing the center of mass to move downward and located at a place close to the case for which diffusion is the only force responsible for the dissolution. On the other hand, for a fixed moderate rotation speed, as $Ra$ increases, the center of mass moves further away from the corresponding no rotation case. Similar to the dissolution of the solute, mixing of the dissolved solute in the solvent depends on the resultant of buoyancy and cylinder rotation. We conclude that there is an optimal pair of $Ra$ and $\Omega$ that causes the dissolved solute to mix better in the solvent. Quantification of this optimal pair is beyond the scope of the current study and will be part of our future research. 

As buoyancy dominates over rotational convection, the initial circular-shaped interface changes to an egg-like shape, with the pointy edge becoming increasingly sharp as $Ra$ increases. However, when rotation dominates over buoyancy, the interface nearly retains its initial circular shape throughout dissolution. A modified Rayleigh number analysis revealed that when $\sqrt{Ra_\Omega}\lesssim 250$, the interface always retains its circular shape for any combination of $Ra (< 10^8)$ and $\Omega$. Although their inclination with the vertical $x$-axis was different, the interface shapes for the same $Ra_\Omega$ obtained through different combinations of $Ra$ and $\Omega$ are seen to exhibit the same characteristics at any instant of dissolution. 

\textcolor{blue}{
Although for dissolution processes, $Sc$ is typically $\mathbf{O}(10^2-10^4)$, here we have restricted our discussion to $Sc = 1$, in accordance with earlier studies in the literature \citep{HUANG, Nandi4}. Nonetheless, additional numerical experiments for $Sc \geq 10$ yielded convergent results ensuring the robustness of our numerical scheme. Detailed quantitative analyses for $Sc$ relevant to some practical dissolution problems will be worth considering in future.} 
As a general proposition, it is a compelling question to examine the relevant effects of rotation on the dissolution of a solute in the contexts of pharmaceuticals, food science, chemical engineering, and environmental science, having applications in drug release into the bloodstream, ingredient dissolution in liquids, metal extraction from ores, and pollutant dispersion in water. The present study paves the way to better understand the role of rotation on convective dissolution.

\vspace{0.2 in}
\textbf{Declaration of interests.} 
The authors report no conflicts of interest. 

\vspace{0.2 in}
\textbf{Authorship contribution statement.}
{\bf Subhankar Nandi:} Conceptualization, Formal analysis, Investigation, Validation, Visualization, Writing – original draft, Writing – review \& editing. {\bf Jiten C. Kalita:} Formal analysis, Validation, Visualization, Writing – review \& editing. {\bf Sanyasiraju VSS Yedida:} Formal analysis, Validation, Writing – review \& editing. {\bf Satyajit Pramanik:} Conceptualization, Formal analysis, Investigation, Validation, Visualization, Writing – original draft, Writing – review \& editing. 

\vspace{0.2 in}
\textbf{Acknowledgments.} 
This work was supported by the Science and Engineering Research Board, Department of Science and Technology, Government of India, through the Core Research Grant (CRG/2023/004156). 
S.P. acknowledges financial support through the Start-Up Research Grant (SRG/2021/001269), MATRICS Grant (MTR/2022/000493) from the Science and Engineering Research Board, Department of Science and Technology, Government of India, and Start-up Research Grant (MATHSUGIITG01371SATP002), IIT Guwahati. 

%
%
%
%
%
%

\appendix
\begin{appendix}

\section{Validations} \label{app:validation}
As a benchmark for validating the developed numerical method, we consider the classical Frank disk problem, which admits a self-similar solution describing the evolution of a moving boundary in an infinite medium. This problem, previously studied by \citet{HUANG} and \citet{GIBOU2}, is a Stefan-type model involving temperature diffusion and interface motion without fluid convection. The temperature field $T(x,t)$ evolves according to

\begin{equation}
\frac{\partial T}{\partial t} = \Delta T \quad \text{in } \Sigma_l(t), \qquad V_n = St \frac{\partial T}{\partial n} \quad \text{on } \Gamma(t),
\end{equation}
with a radially symmetric interface whose exact position is given by $R(t) = S_0 \sqrt{t}$, and the corresponding temperature distribution is
\begin{equation}
    T(x,t) = 
\begin{cases}
0, & s \leq S_0, \\[4pt]
T_\infty \left(1 - \dfrac{F(s)}{F(S_0)}\right), & s > S_0,
\end{cases}
\end{equation}
where $s = \dfrac{|x|}{\sqrt{t}}$, $F(s) = E_1\left( \dfrac{s^2}{4} \right)$ and $E_1(z) = \int_z^\infty \frac{e^{-t}}{t} dt$ is the exponential integral. Following \citet{HUANG}, we set $St = -0.4$, $S_0 = 1.2$, initial time of the numerical simulation to be $t_0 = 0.1$, and an initial interface radius $R(0.1) = 0.38$. The far-field temperature $T_\infty$ is computed from the Stefan condition using the exact solution, yielding $T_\infty \approx -0.999$. The computational domain is chosen sufficiently large so that the moving interface remains well within the domain throughout the simulation up to the final time $t = 1$.

\begin{figure}
  \centering
  \begin{subfigure}[b]{0.315\textwidth}
    \includegraphics[width=\textwidth]{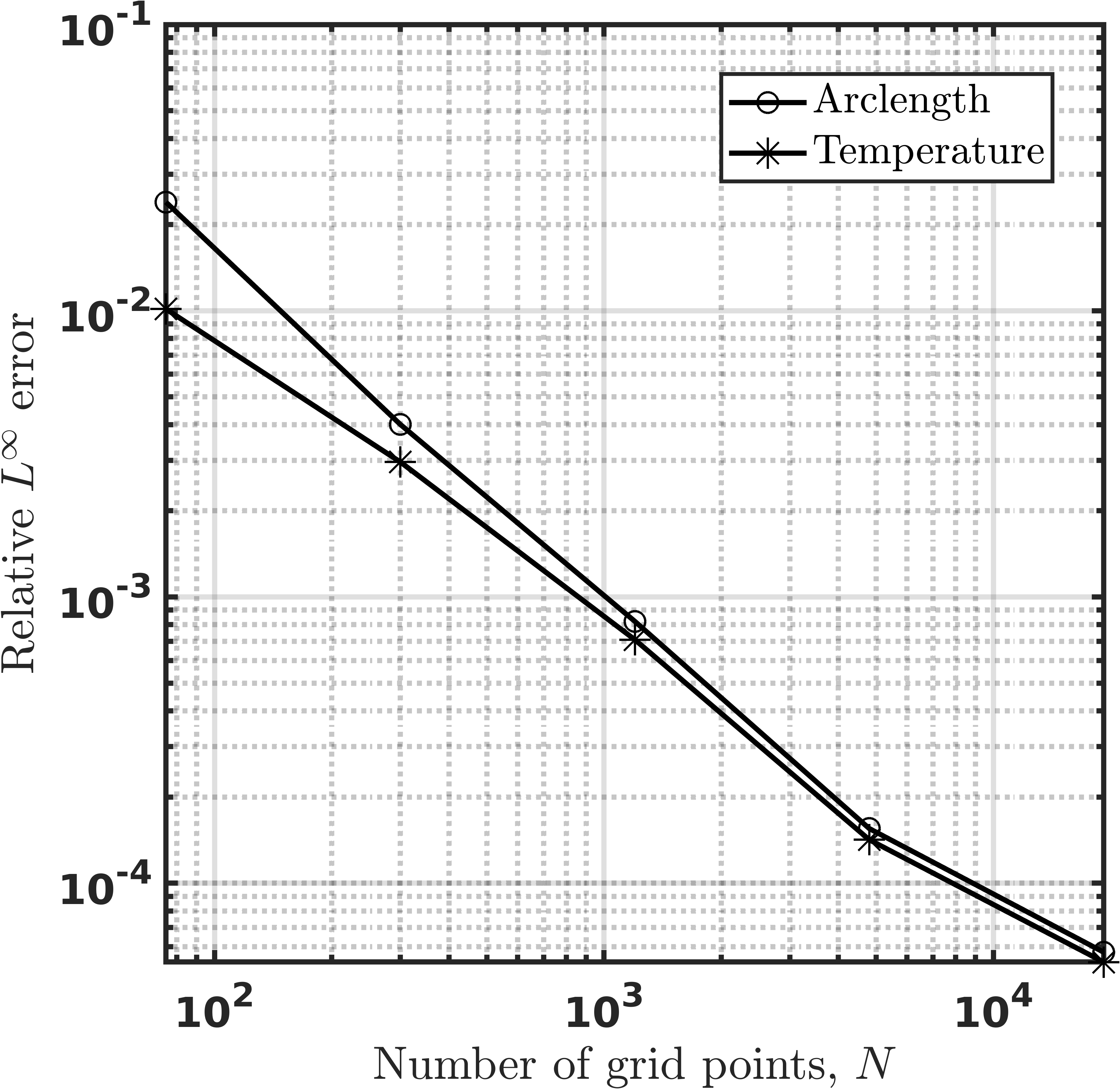}
    \caption{}\label{fig:conver}
  \end{subfigure}
  \hfill
  \begin{subfigure}[b]{0.325\textwidth}
    \includegraphics[width=\textwidth]{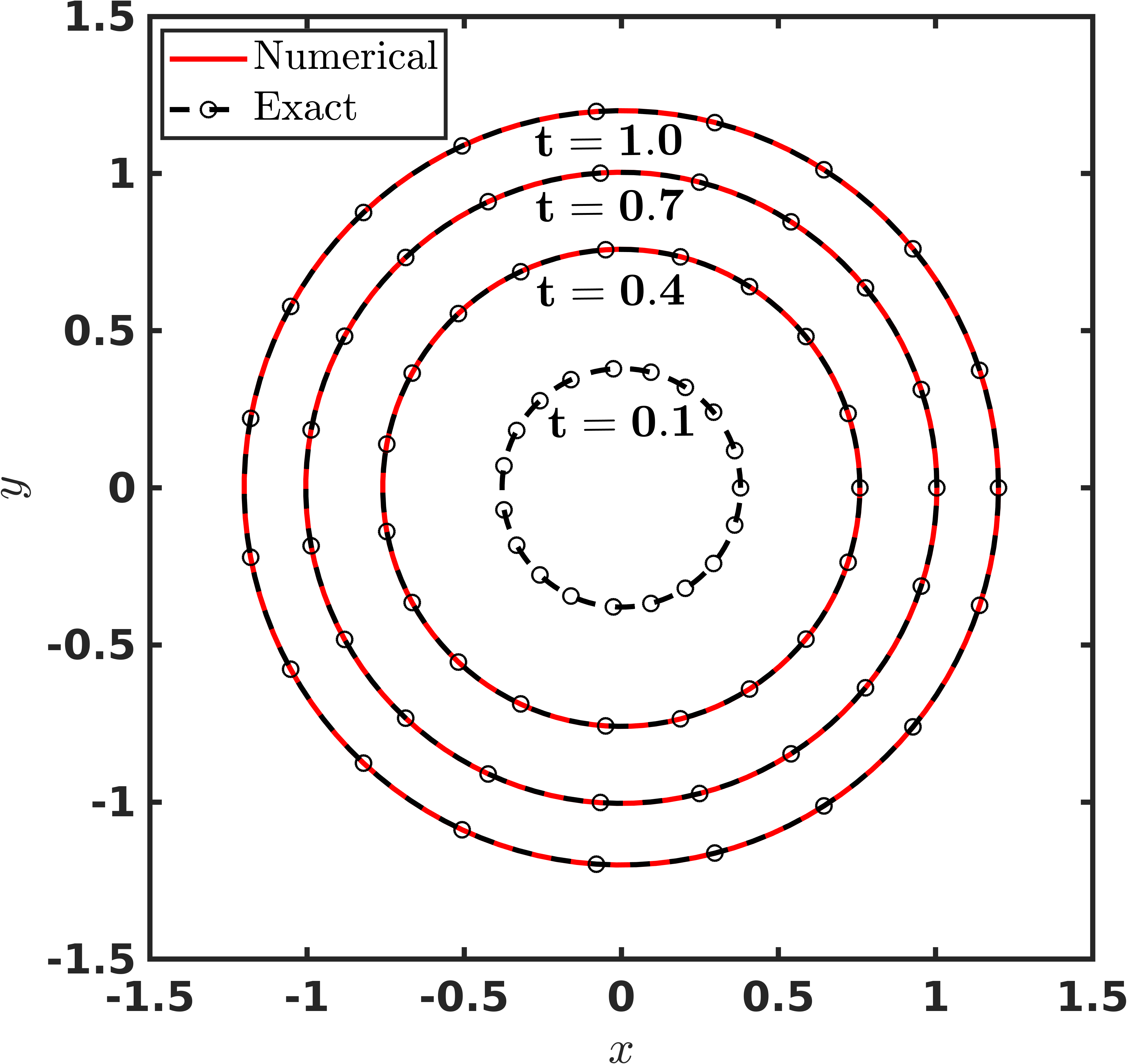}
    \caption{}\label{fig:inter}
  \end{subfigure}
  \hfill
  \begin{subfigure}[b]{0.30\textwidth}
    \includegraphics[width=\textwidth]{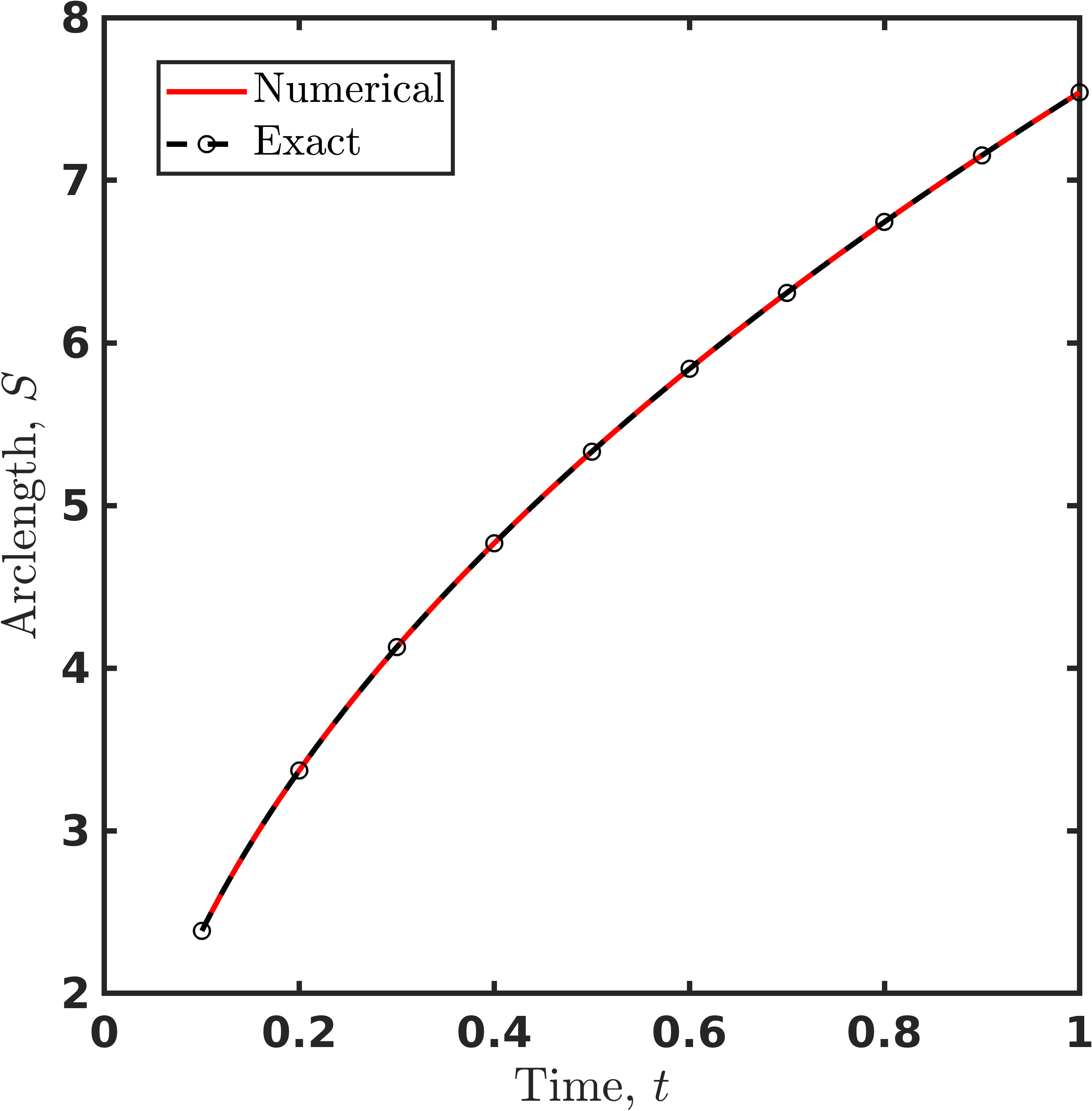}
    \caption{}\label{fig:Arc}
  \end{subfigure}
\caption{(a) Grid convergence analysis showing the relative $L^\infty$ errors in arclength and temperature with respect to the number of grid points, where the error norm is computed over all time steps. (b) Evolution of the interface at four time instances: $t = 0.1$, $0.4$, $0.7$, and $1.0$, where the initial shape corresponds to $t = 0.1$. In each case, the numerical solution is shown using solid red curves and the exact interface is indicated by black dashed curves with circular markers. (c) Comparison (following the same graphical representation as (b)) of the total arclength of the interface over time from numerical simulation and the exact solution.}
\label{fig:problem1}
\end{figure}

We provide a grid convergence analysis in Figure \ref{fig:problem1}(a), where five different grid configurations ($5\times15$, $10\times30$, $20\times60$, $40\times120$, and $80\times240$) are used to compute the relative $L^\infty$ errors in temperature and interface arclength. The errors are evaluated over the entire simulation time. The results indicate the accuracy of our simulation, and a grid of size $40\times120$ was found to be large enough to accurately capture the phenomenon. As such, we have used this grid setup in all our subsequent simulations. Figure \ref{fig:problem1}(b) illustrates the evolution of the moving interface at four different times ($t = 0.1$, $0.4$, $0.7$, and $1.0$), with the numerical and exact interfaces in close agreement. Figure \ref{fig:problem1}(c) depicts that the time evolution of the arclength of the interface computed numerically is indistinguishable from that computed using the the exact solution, indicating the accuracy of the proposed method. 

Next, we consider a configuration similar to our primary study to further validate the numerical method. Specifically, we examine natural convective heat transfer between concentric horizontal cylinders, where the inner cylinder is heated and the annular region contains cold liquid, in the absence of moving boundaries. We first focus on the case without rotation, which has been extensively studied in the literature through experimental \citep{KUEHN, BISHOP}, numerical \citep{KUEHN, SARKAR, INGHAM}, and semi-analytical \citep{mack1968natural, ABBOTT} approaches. 

\begin{table}
\begin{center}
\begin{tabular}{l l cc}
 \thead{Grid \\size}   & $\psi_{min}$ & \thead{ $Nu$ at \\inner wall} & \thead{$Nu$ at \\outer wall}  \\
$6 \times 21$  & -4.1624 & 1.1215 & 1.1264  \\
$11 \times 41$  & -4.2199 & 1.1037 & 1.1130  \\
$21 \times 81$   & -4.2999 & 1.1101 & 1.1112 \\
$41 \times 161$  & -4.3664 & 1.1167 & 1.1164  \\
$81 \times 321$ & -4.3695 & 1.1173 & 1.1182 \\
\end{tabular}
\caption{Grid independence study for $Ra = 3000$, $Pr = 0.7$, and radius ratio 1.85. The minimum value of stream function ($\psi_{min}$) and average Nusselt numbers at the inner and outer walls are shown for different grid resolutions. Results converge toward the theoretical value $\psi_{min} = -4.36$ reported by \citet{mack1968natural}.} 
\label{table:mack}
\end{center}
\end{table}

We perform a grid convergence study by comparing our numerical results with the theoretical findings of \citet{mack1968natural}, which employed a power series method to solve the problem. In their approach, both the temperature and stream function were expanded in powers of the Rayleigh number. Using the first three terms of the series, the stream function value at the stagnation point was reported as $-4.36$ for $Ra = 3000$, $Pr = 0.7$, and a radius ratio $1.85$. We compute the minimum streamfunction value $\psi_{\min}$ for several grid resolutions, as summarized in Table \ref{table:mack}. The results ensures convergence of the numerical results to the theoretical value, confirming the grid independence of the solution. Additionally, the average Nusselt numbers ($Nu$) on both the inner and outer walls exhibited convergence across the tested grid resolutions, as shown in Table \ref{table:mack}. Based on this, we adopt a grid resolution of $41 \times 161$ for subsequent simulations. 

\begin{table}
\begin{center}
\begin{tabular}{l cc cc cc}
 & \multicolumn{2}{c}{\thead{Present }} & \multicolumn{2}{c}{\thead{ \citet{KUEHN} } } & \multicolumn{2}{c}{\thead{\citet{SARKAR} } } \\
\cmidrule(lr){2-3} \cmidrule(lr){4-5} \cmidrule(lr){6-7}
 $Ra$     & \thead{Inner} & \thead{Outer}  & \thead{Inner } & \thead{ Outer}  & \thead{Inner} & \thead{ Outer}  \\
$10^2$  & 1.0009 & 1.0009 & 1.000  & 1.002 & 1.001  & 1.001  \\
$10^3$  & 1.0778 & 1.0775 & 1.081  & 1.084 & 1.082  & 1.082 \\
$10^4$   & 1.9781 & 1.9654 & 2.010  & 2.005 & 1.989  & 1.989 \\
$5 \times 10^4$  & 2.9652 & 2.9540 & 3.024  & 2.973  & 2.961  & 2.961\\
\end{tabular}
\caption{Comparison of the computed average Nusselt number ($Nu$) with available numerical results from the literature \citep{KUEHN, SARKAR} for fixed parameters values, $L/D_i = 0.8$, $Pr = 0.7$, and various Rayleigh numbers ($Ra$).} 
\label{table:kuehn}
\end{center}
\end{table}

\begin{figure}
 \begin{center}
 \includegraphics[scale=0.82]{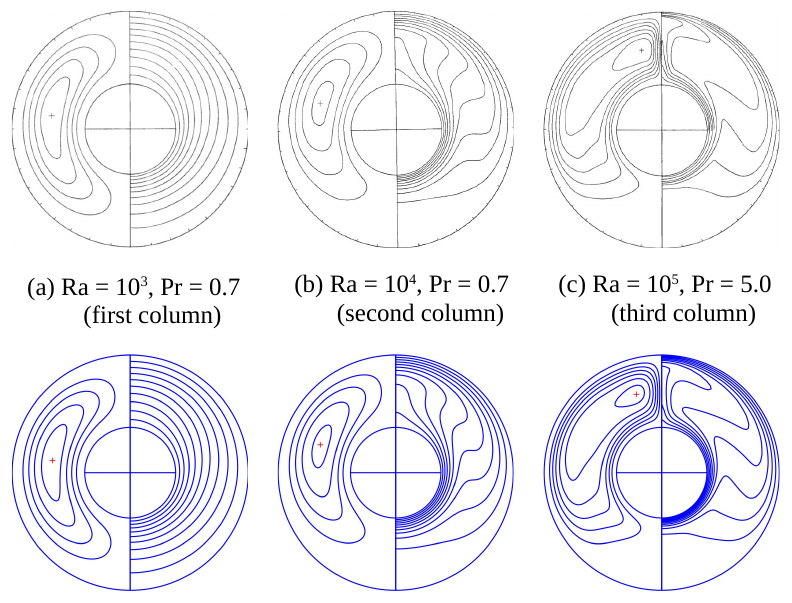}
  \caption{The qualitative comparison of the streamlines and isotherms obtained using the present method (second row) with the published results \citep{KUEHN} (first row) for fixed $L/Di = 0.8$.} \label{figg1}
\end{center} 
\end{figure}

The comparison of average Nusselt number values presented in Table \ref{table:kuehn} demonstrates excellent agreement between the present results and those available in the literature \citep{KUEHN, SARKAR}. This confirms that the chosen grid resolution remains sufficiently accurate even at relatively high Rayleigh numbers. Furthermore, a qualitative comparison of streamlines and isotherms obtained using the present method (second row) with the published results of \citet{KUEHN}, as shown in Figure \ref{figg1}, also revealed an excellent agreement. 
  
We now extend the validation by considering the same annular configuration with rotational effects, which makes the setup even closer to the problem considered in this study. Here, we consider the problem of heat transfer between two counter-rotating concentric cylinders for which both numerical and experimental results are available in the existing literature \citep{abu2007combined, singh1982heat, lee1992numerical, fu1994enhancement, launder1974laminar}. The counter-rotation is achieved by rotating the inner cylinder counterclockwise with a constant angular speed $\Omega_i$, while the outer cylinder rotates in the clockwise direction with an angular speed $\Omega_o$.

\begin{table}
\begin{center}
\begin{tabular}{l cc cc }
\toprule
 & \multicolumn{2}{c}{\thead{Present Study}} & \multicolumn{2}{c}{\thead{\citet{abu2007combined} } }  \\
\cmidrule(lr){2-3} \cmidrule(lr){4-5} 
 $\delta$     & \thead{$\psi_{min}$} & \thead{$\psi_{max}$}  & \thead{$\psi_{min}$ } & \thead{ $\psi_{max}$}   \\
\midrule
$0.1$  & -0.2583 & 0.0181 & -0.2680  & 0   \\
$0.5$  & -0.2885 & 0 & -0.2920  & 0  \\
$1.0$   & -0.3338 & 0 & -0.3220  & 0.019 \\
\bottomrule
\end{tabular}
\end{center}
\caption{Comparison of $\psi_{min}$ and $\psi_{max}$ values obtained using the present method with the existing numerical results \citep{abu2007combined} for various values of $\delta$.} 
\label{table:abu}
\end{table}

Computations are performed using $41 \times 161$ spatial grid points, which is chosen based on a grid refinement study that minimizes the error in the minimum $(\psi_{min})$ and maximum $(\psi_{max})$ values of the stream function. We compare the present numerical results with those of \citet{abu2007combined} for $R (:= r_o/r_i) = 2$, $Re = 150$, $Ra = 10^4$, $Pr = 0.7$, and $\delta (:= \Omega_o r_o / \Omega_i r_i) \in \{ 0.5, 1 \}$ in Table \ref{table:abu}, which shows a consistent trend with the existing results.

\begin{figure}
 \begin{center}
 \includegraphics[width=\textwidth]{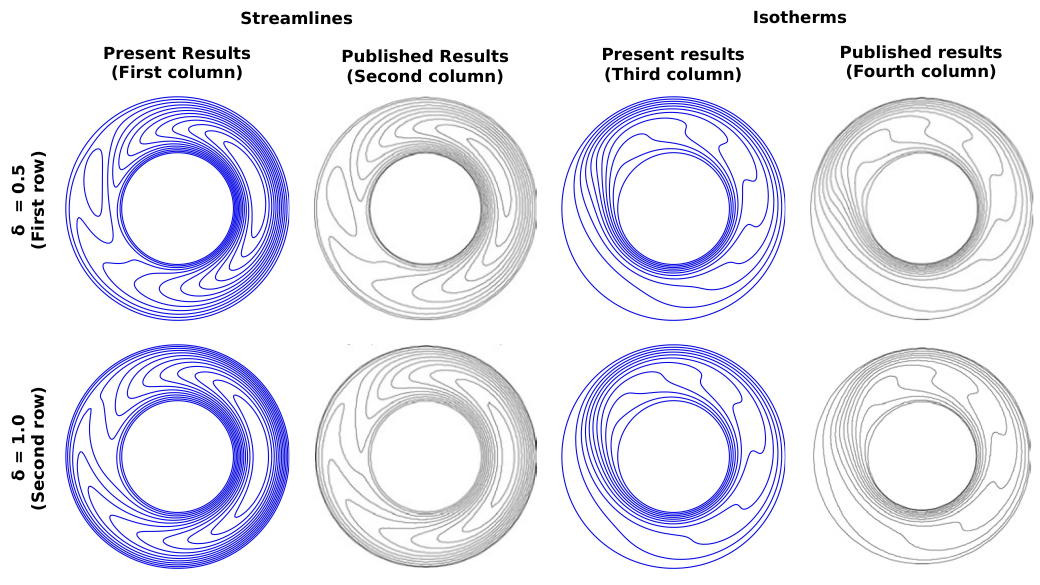}
  \caption{A qualitative comparison of the streamlines (first and second columns) and isotherms (third and fourth columns) obtained in the present study (first and third columns, respectively) with the results from \citet{abu2007combined} (second and fourth columns, respectively) for velocity ratios of $\delta = 0.5, 1.0 $, with a fixed radius ratio of $R = 2$.} 
  \label{fig:validation}
\end{center} 
\end{figure}
Figure \ref{fig:validation} demonstrates that the present method accurately captured the flow patterns and temperature distribution, depicting a good agreement with the results of \citet{abu2007combined}. 

\section{Grid independence} \label{app:grid_independence}
This study considers six distinct sets of spatial grid points: $10 \times 60$, $20 \times 120$, $30 \times 180$, $40 \times 240$, $50 \times 300$, and $60 \times 360$, with a fixed time step $\Delta t = 10^{-3}$ for the numerical experiments. In each case, the solutions are computed for $Ra = 10^6$, $10^7$, $10^8$, and $\Omega = 1$ at two time instances — $t = 10$ (developing flow) and $t = 20$ (fully developed flow) — to capture the flow effects induced by the phase change.

The grid independence is tested by comparing the percentage errors for the average Sherwood number $Sh(t)$, dissolved solute area $A_d(t)$, and the maximum value of the stream function $\psi_{\max}(t)$ between two successive grid point sets. The average Sherwood number representing the overall mass transfer rate along the interface boundary is defined as
\begin{equation}
\label{eq:sh_nondim}
Sh(t) = \frac{1}{S(t)} \int_{0}^{S(t)} Sh^\prime(t) ds,
\end{equation}
where $Sh^\prime(t)$ is the non-dimensional local Sherwood number for mass transfer rate given by
\begin{equation}
\label{eq:sh_nondim_local}
Sh^\prime(t) = - \frac{\partial c}{\partial n^\Gamma},
\end{equation}
with $S(t)$ being the arc length of the interface. The dissolved area fraction over time is quantified by
\begin{equation}
\label{eq:Ad}
A_d(t) = 1 - \frac{A_s(t)}{A_s(0)},
\end{equation}
where $A_s(t)$ is the area of the remaining (undissolved) solute at time $t$.


Relative percentage error for each quantity is calculated with respect to the value obtained on the next finer grid, using
\begin{equation}
\text{Error}(\%) = \frac{\left| Q_{\text{finer}} - Q_{\text{current}} \right|}{\left| Q_{\text{current}} \right|} \times 100 ,
\end{equation}
where $Q$ denotes the computed value of $Sh$, $\psi_{\max}$, or $A_d(t)$. 

The relative percentage error in $Sh(t)$, $A_d(t)$, and $\psi_{\max}(t)$ at $t = 10$ and $t = 20$ for $\Omega = 1$ are computed corresponding to different grid resolutions for $Ra$ varying two-order of magnitudes to ensure the robustness of the numerical scheme. For $Ra = 10^6$, the relative percentage error between the two finest grid sets are approximately 0.92\% for $Sh(t)$, 0.21\% for $A_d(t)$, and 0.55\% for $\psi_{\max}(t)$ at $t = 10$ (see Table \ref{tab:refinement_all}). At $t = 20$, these errors are 1.72\%, 0.57\%, and 0.57\%, respectively (see Table \ref{tab:refinement_all}). 
As evident from Table \ref{tab:refinement_all}, the relative percentage errors in $Sh(t)$, $A_d(t)$, and $\psi_{\max}(t)$ at both $t = 10$ and $t = 20$ remain consistently below 2.5\% for the grid size $50 \times 300$. Therefore, we use the spatial resolution of $50 \times 300$, and a time step size $\Delta t = 10^{-3}$ in all the simulations discussed in this paper. 


\begin{table}
\centering
\begin{tabular}{l l l l r l r l r}
$Ra$ & $t$ & Grid Size & $Sh$ & Err (\%) & $\psi_{\max}$ & Err (\%) & $A_d$ & Err (\%) \vspace{0.3 cm} \\
$10^{6}$
 & 10 & $10\times60$ & 4.1111 & 16.11\% & 0.2995 & 14.36\% & 0.3815 & 9.86\% \\
 &  & $20\times120$ & 3.4489 & 2.56\% & 0.2565 & 1.72\% & 0.3439 & 2.97\% \\
 &  & $30\times180$ & 3.3605 & 0.98\% & 0.2521 & 0.56\% & 0.3337 & 0.30\% \\
 &  & $40\times240$ & 3.3935 & 1.16\% & 0.2535 & 0.75\% & 0.3327 & 0.18\% \\
 &  & $50\times300$ & 3.4327 & 0.92\% & 0.2554 & 0.55\% & 0.3333 & 0.21\% \\
 &  & $60\times360$ & 3.4644 & --     & 0.2568 & --     & 0.3340 & --   \vspace{0.2 cm} \\
 & 20 & $10\times60$ & 2.6581 & 4.15\% & 0.4098 & 20.60\% & 0.5874 & 7.03\% \\
 &  & $20\times120$ & 2.7685 & 2.38\% & 0.3254 & 14.51\% & 0.5461 & 1.61\% \\
 &  & $30\times180$ & 2.8344 & 2.43\% & 0.2782 & 3.81\% & 0.5373 & 0.47\% \\
 &  & $40\times240$ & 2.9033 & 2.21\% & 0.2676 & 1.38\% & 0.5398 & 0.67\% \\
 &  & $50\times300$ & 2.9674 & 1.72\% & 0.2639 & 0.57\% & 0.5434 & 0.57\% \\
 &  & $60\times360$ & 3.0183 & --     & 0.2624 & --     & 0.5465 & --    \vspace{0.3 cm} \\
 $10^{7}$ 
 & 10 & $10\times60$ & 4.7886 & 23.87\% & 0.2457 & 8.73\% & 0.4159 & 14.85\% \\
 &  & $20\times120$ & 3.6317 & 17.56\% & 0.2244 & 5.77\% & 0.3538 & 12.20\% \\
 &  & $30\times180$ & 2.9896 & 3.85\% & 0.2115 & 1.89\% & 0.3090 & 5.44\% \\
 &  & $40\times240$ & 2.8850 & 0.88\% & 0.2095 & 0.95\% & 0.2967 & 0.40\% \\
 &  & $50\times300$ & 2.9109 & 0.92\% & 0.2082 & 0.62\% & 0.2944 & 0.24\% \\
 &  & $60\times360$ & 2.9382 & --     & 0.2074 & --     & 0.2949 & --    \vspace{0.2 cm} \\
& 20 & $10\times60$ & 3.7100 & 7.19\% & 0.3656 & 45.63\% & 0.6636 & 13.61\% \\
 &  & $20\times120$ & 3.4453 & 15.91\% & 0.1979 & 15.54\% & 0.5730 & 11.16\% \\
 &  & $30\times180$ & 2.9146 & 5.62\% & 0.1671 & 3.48\% & 0.5094 & 2.91\% \\
 &  & $40\times240$ & 2.8201 & 1.24\% & 0.1614 & 1.05\% & 0.4946 & 0.41\% \\
 &  & $50\times300$ & 2.8556 & 1.22\% & 0.1586 & 0.97\% & 0.4956 & 0.40\% \\
 &  & $60\times360$ & 2.9132 & --     & 0.1541 &  --    & 0.4979 &  --   \vspace{0.3 cm} \\
  $10^{8}$ 
 & 10 & $10\times60$ & 5.0264 & 23.87\% & 0.2335 & 10.24\% & 0.4284 & 16.36\% \\
 &  & $20\times120$ & 3.8268 & 30.36\% & 0.2096 & 13.07\% & 0.3583 & 21.35\% \\
 &  & $30\times180$ & 2.6651 & 17.76\% & 0.1822 & 6.26\% & 0.2818 & 13.66\% \\
 &  & $40\times240$ & 2.1918 & 7.47\% & 0.1708 & 2.28\% & 0.2433 & 8.51\% \\
 &  & $50\times300$ & 2.0281 & 1.73\% & 0.1669 & 1.92\% & 0.2226 & 0.85\% \\
 &  & $60\times360$ & 1.9930 & --     & 0.1637 & --     & 0.2207 & --  \vspace{0.2 cm} \\
 & 20 & $10\times60$ & 4.2597 & 7.19\% & 0.2552 & 24.02\% & 0.6906 & 14.84\% \\
 &  & $20\times120$ & 3.9533 & 27.54\% & 0.1939 & 15.52\% & 0.5881 & 20.27\% \\
 &  & $30\times180$ & 2.8645 & 20.62\% & 0.1638 & 11.42\% & 0.4689 & 12.82\% \\
 &  & $40\times240$ & 2.2739 & 9.24\% & 0.1451 & 7.10\% & 0.4088 & 5.99\% \\
 &  & $50\times300$ & 2.0639 & 2.46\% & 0.1348 & 0.82\% & 0.3843 & 1.35\% \\
 &  & $60\times360$ & 2.0131 & --     & 0.1337 & --     & 0.3791 & --  \\
\end{tabular}
\caption{Relative percentage error corresponding to different grid sizes for $Sh$, $\psi_{\max}$, and $A_d(t)$ for different values of $Ra$, $\Omega = 1$ at $t = 10$ and $20$.}
\label{tab:refinement_all}
\end{table}

\begin{figure}
  \centering
  \begin{subfigure}[b]{0.32\textwidth}
    \includegraphics[width=\textwidth]{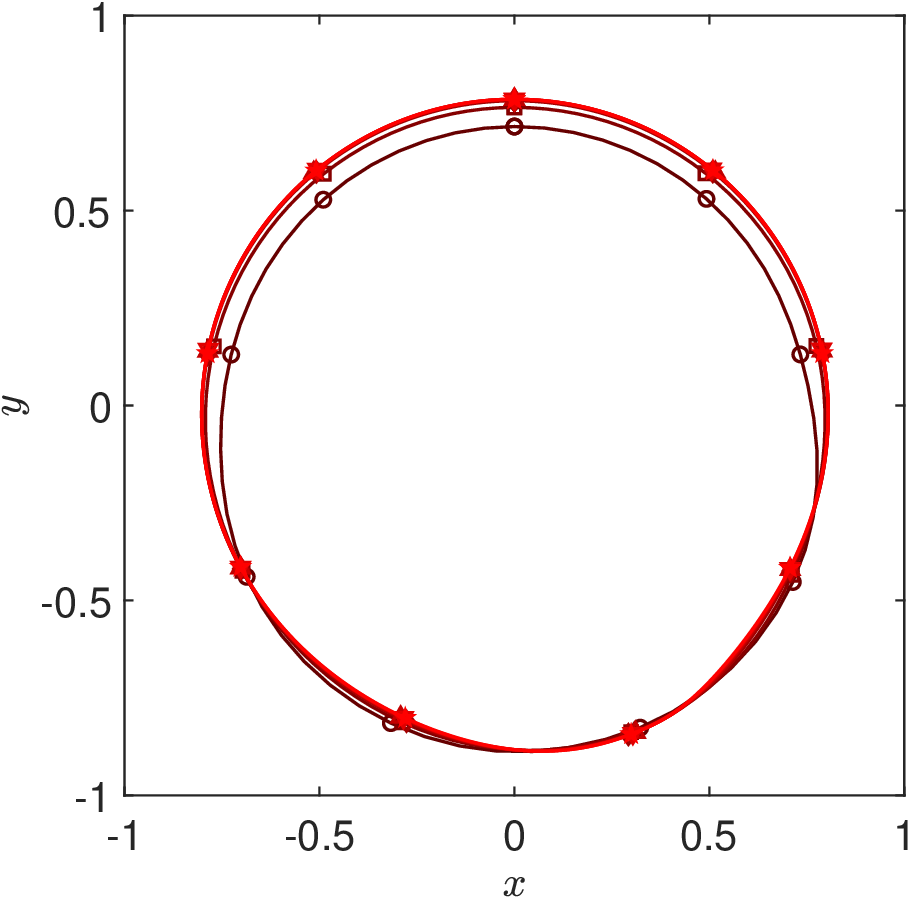}
    \caption{}\label{fig:shape10}
  \end{subfigure}
  \hfill
  \begin{subfigure}[b]{0.32\textwidth}
    \includegraphics[width=\textwidth]{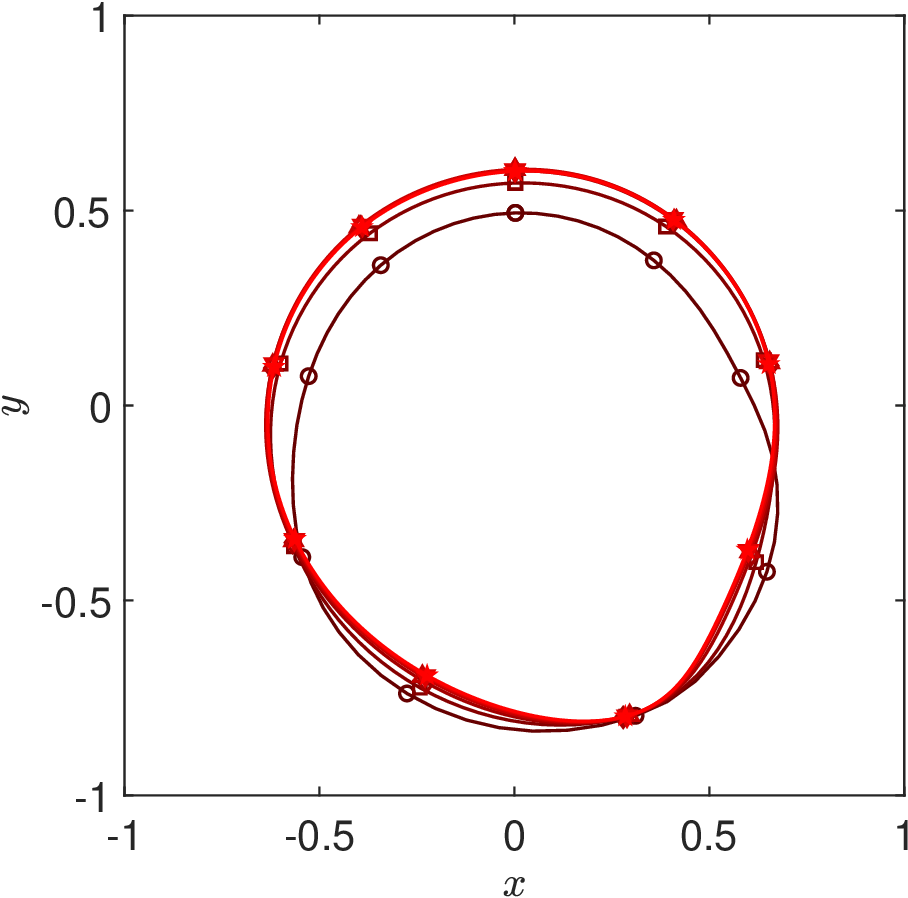}
    \caption{}\label{fig:shape20}
  \end{subfigure}
  \hfill
  \begin{subfigure}[b]{0.32\textwidth}
    \includegraphics[width=\textwidth]{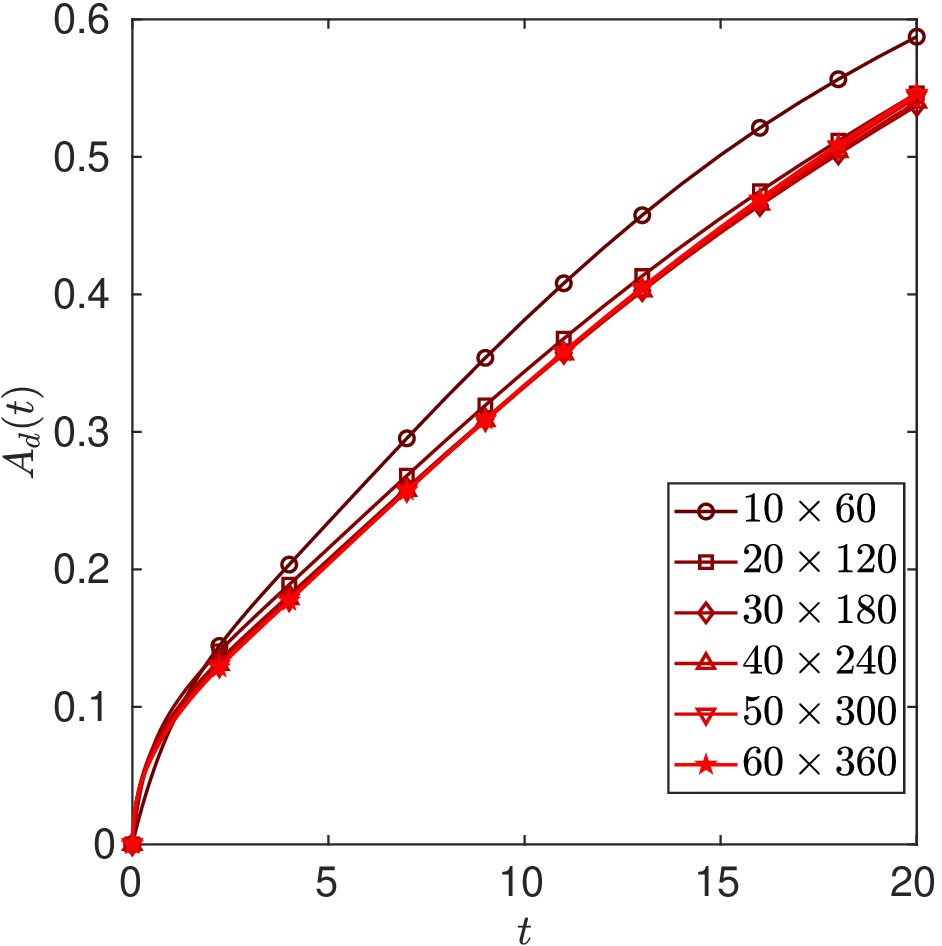}
    \caption{}\label{fig:Areatime}
  \end{subfigure}
  \caption{Grid refinement study of the dissolution solver: (a, b) Instantaneous shape comparison at $t = 10$ and $t = 20$; (c) Evolution of dissolved area $A_d(t)$ up to $t = 20$.}
  \label{fig:grid_independence1}
\end{figure}

To further strengthen the grid refinement assessment, we also include direct comparisons of the instantaneous interface shapes and area evolution. Figure \ref{fig:grid_independence1}(a–b) presents the interface shapes at $t = 10$ and $t = 20$ for different grid resolutions. The results demonstrate excellent agreement among the three finest grid sets, especially the $40 \times 240$, $50 \times 300$, and $60 \times 360$ cases. Figure \ref{fig:grid_independence1}(c) shows the time evolution of the dissolved area $A_d(t)$ up to $t = 20$ for all grids, with the three finest grids converging closely throughout the entire period. 

\end{appendix}





\end{document}